# *Bounds on the Topological Charge of Photonic Systems*

**Filipa R. Prudêncio[1,2*], I. Souza[3,4] and Mário G. Silveirinha[1]**

[1]University of Lisbon – Instituto Superior Técnico and Instituto de Telecomunicações, Avenida Rovisco Pais 1, 1049-001 Lisbon, Portugal

[2]Instituto Universitário de Lisboa (ISCTE-IUL), Avenida das Forças Armadas 376, 1600-077 Lisbon, Portugal

[3]Centro de Física de Materiales, Universidad del País Vasco, 20018 San Sebastián, Spain

[4]Ikerbasque Foundation, 48013 Bilbao, Spain

## Abstract

Topology has become a central concept in understanding physical phenomena, leading to important advances in condensed matter and photonics. Recent work has established a universal upper bound on the energy gap of Chern insulators in electronic systems, revealing a fundamental connection between topology, quantum geometry, and optical absorption. Here, we generalize this framework to photonic systems, deriving rigorous upper bounds on gap Chern numbers without requiring explicit topological analysis. Our approach enables the estimation of the topological charge of band gaps in both dispersive and nondispersive regimes.

[*] Corresponding author: filipa.prudencio@lx.it.pt

# I. INTRODUCTION

The role of topology in determining physical responses has become a central theme in contemporary condensed-matter and photonic research [1-11]. The topological nature of a physical system is usually determined by the global properties of the governing operator, commonly the Hamiltonian in quantum systems. Among the most studied examples are Chern insulators, which break time-reversal symmetry and are characterized by an integer topological invariant, the Chern number [3, 4, 8]. Their non-trivial topology gives rise to unidirectional, gapless edge states that are protected against scattering, leading to transport and optical responses that remain robust under disorder, imperfections, and other perturbations [10].

A defining feature of insulating phases of matter is the presence of an energy gap separating the ground state from excited states. In the last few decades, it was revealed that Chern insulators exhibit properties that are fundamentally distinct from those of trivial insulators, reflected for example in a quantized Hall conductivity [1]. Furthermore, recent theoretical advances have established universal bounds on the energy gap of Chern insulators, uncovering a direct connection between topological invariants, quantum geometry, and optical response [12, 13, 14]. This perspective builds on earlier geometric and optical-sum-rule formulations of insulating matter, where optical response is related to electronic localization, the quantum metric [14], and magnetic circular dichroism [15-17]. Related bounds connect topology and quantum geometry to structure factors [18, 19], while broader optical sum-rule constraints relate gaps, dielectric response, and localization in insulating matter [20, 21].

While such bounds are now well established in electronic systems, extending them to photonics requires additional considerations. In electronic settings, the relevant bounds are naturally expressed in terms of quantities such as effective mass and carrier density [12], which do not have immediate analogues in the optics context. Moreover, the intrinsically dispersive character of most photonic materials further complicates any straightforward translation of electronic results.

In this work, we derive a general class of bounds for both nondispersive and dispersive photonic systems. In the nondispersive case, the variation of the optical material parameters with frequency is neglected. The separate treatment of the two cases stems from the different structure of the governing operators: whereas for nondispersive systems the relevant Bloch operator is often quadratic in the wave vector **k**, analogous to the Hamiltonian of electronic systems, when dispersion is considered the photonic Bloch operator is generally linear in **k**, reflecting the inherently relativistic nature of light. This difference leads to a qualitatively distinct class of bounds. In both nonrelativistic electronic systems and nondispersive photonic systems, the gap Chern number bound is controlled by the curvature of the Hamiltonian. By contrast, in relativistic-type photonic systems, the bound emerges directly from the fundamental constraint imposed by the speed of light in the medium. These velocity bounds are found to be relatively loose in the examples we consider. To address this limitation, we develop a strategy that allows the nondispersive bounds to be applied as well to dispersive systems in a practically useful way.

The article is organized as follows. Section II derives upper bounds on gap Chern numbers, treating separately systems described by operators that are quadratic and linear in the wave vector. In particular, we derive a new velocity bound for electromagnetic platforms, explain its relativistic origin, and discuss its extension to non-Hermitian systems. Furthermore, we introduce a curvature bound that generalizes the Onishi-Fu result [12]. In Sec. III, we illustrate the application of the curvature bound to photonic systems, using two representative platforms: magnetized gyrotropic photonic crystals and plasmonic photonic crystals with Tellegen-type magnetoelectric coupling. Section IV applies the velocity bound to dispersive electromagnetic systems. We discuss the challenges that arise in this context and introduce a strategy that enables the curvature bounds to be used effectively in dispersive settings. Finally, Sec. V summarizes the main results and conclusions.

## II. BOUNDS ON THE CHERN NUMBER

We consider a family of Hermitian operators $\hat{H}_{\mathbf{k}}$ parameterized by a two-component vector $\mathbf{k}$. Typically, $\hat{H}_{\mathbf{k}}$ governs the wave dynamics of the physical system under consideration, whereas $\mathbf{k}$ represents a wave vector in the Brillouin zone. For simplicity, we shall assume that $\hat{H}_{\mathbf{k}}$ is periodic and smooth in $\mathbf{k}$. In some tight-binding conventions, as well as in continuum Bloch formulations, $\hat{H}_{\mathbf{k}+\mathbf{G}}$ may be unitarily equivalent to $\hat{H}_{\mathbf{k}}$, rather than strictly identical to it. This refinement does not affect the analysis below. The theory applies equally to systems with fermionic or bosonic excitations.

The spectral problem under consideration is:

$$\hat{H}_{\mathbf{k}} \left| u_{n\mathbf{k}} \right\rangle = E_{n\mathbf{k}} \left| u_{n\mathbf{k}} \right\rangle , \tag{1}$$

where $E_{n\mathbf{k}}$ are the eigenvalues of the operator $\hat{H}_{\mathbf{k}}$ , given as a function of the wavevector $\mathbf{k}$, and $\left| u_{n\mathbf{k}} \right\rangle$ are the corresponding eigenfunctions (*n* labels the bands). When a set of bands is separated from the rest of the spectrum by a finite gap, the operator family $\hat{H}_{\mathbf{k}}$ may have a nontrivial topological structure characterized by a Chern number. It is well known that a nonzero Chern number is only possible when time-reversal symmetry is broken. Such nontrivial topological charge can have remarkable physical consequences, most notably the emergence of unidirectional edge states at the system boundary. In the following, building on Ref. [12], we show that the topological charge is fundamentally constrained by very general properties of the operator $\hat{H}_{\mathbf{k}}$. In particular, the gap Chern number is bounded by universal features such as the size of the Brillouin zone, the width of the spectral gap, and global properties of the material system response. The bound does not depend explicitly on geometrical details of the system, such as the shape or size of the material inclusions.

Our starting point is the formula for the Chern number of a subset of bands denoted by $F$ (the "filled bands"). Specifically, the Chern number can be expressed in terms of the modes as [3]:

$$\mathcal{C}_F = \frac{i}{2\pi}\int d^2\mathbf{k} \sum_{\substack{n\mathbf{k}\in F \\ m\mathbf{k}\in E}} \left[ \left\langle u_{n\mathbf{k}} \left| \hat{v}_{\mathbf{k},1} \right| u_{m\mathbf{k}} \right\rangle \left\langle u_{m\mathbf{k}} \left| \hat{v}_{\mathbf{k},2} \right| u_{n\mathbf{k}} \right\rangle - \left\langle u_{m\mathbf{k}} \left| \hat{v}_{\mathbf{k},1} \right| u_{n\mathbf{k}} \right\rangle \left\langle u_{n\mathbf{k}} \left| \hat{v}_{\mathbf{k},2} \right| u_{m\mathbf{k}} \right\rangle \right] \frac{1}{\left[ E_{mn,\mathbf{k}} \right]^2}, \quad (2)$$

where $E_{mn,\mathbf{k}} = E_{m\mathbf{k}} - E_{n\mathbf{k}}$, $\hat{\mathbf{v}}_{\mathbf{k}} = \partial_{\mathbf{k}} \hat{H}_{\mathbf{k}}$ and $E$ denotes the set of bands complementary to $F$ (set of "empty bands"). When $\hat{H}_{\mathbf{k}}$ has units of frequency, the operator $\hat{\mathbf{v}}_{\mathbf{k}}$ can be understood as the velocity. Note that the components of the velocity are $\hat{v}_{\mathbf{k},i} = \partial_i \hat{H}_{\mathbf{k}} \equiv \frac{\partial \hat{H}_{\mathbf{k}}}{\partial k_i}$, $i$=1,2.

Next, we discuss two independent bounds for $\mathcal{C}_F$, whose relevance depends on the specific structure of $\hat{H}_{\mathbf{k}}$.

### *A. Curvature bound*

To begin with, we derive a bound relying on the "curvature" of $\hat{H}_{\mathbf{k}}$, i.e., on the second order derivatives of $\hat{H}_{\mathbf{k}}$ with respect to the momentum: $\nabla_{\mathbf{k}}^2 \hat{H}_{\mathbf{k}} \equiv \frac{\partial^2 \hat{H}_{\mathbf{k}}}{\partial k_1^2} + \frac{\partial^2 \hat{H}_{\mathbf{k}}}{\partial k_2^2}$. Here, "curvature" does not refer to the Berry curvature, but instead to the operator obtained from $\hat{H}_{\mathbf{k}}$ by taking second derivatives with respect to the components of the wave vector, namely the Laplacian in $\mathbf{k}$ space.

To this end, we introduce a non-negative Hermitian tensor $\mathbf{D} = \left[ d_{ij} \right]_{i,j=1,2}$, known as the quantum-geometric tensor, such that:

$$\mathbf{D} = \frac{1}{2\pi} \int d^2\mathbf{k} \sum_{\substack{n\mathbf{k}\in F \\ m\mathbf{k}\in E}} \langle u_{n\mathbf{k}} | \hat{\mathbf{v}}_{\mathbf{k}} | u_{m\mathbf{k}} \rangle \otimes \langle u_{m\mathbf{k}} | \hat{\mathbf{v}}_{\mathbf{k}} | u_{n\mathbf{k}} \rangle \frac{1}{\left[ E_{mn,\mathbf{k}} \right]^2} . \tag{3}$$

Here, $\otimes$ denotes the tensor product of two vectors. It is relevant to note that for a generic vector $\mathbf{w}$, we have

$$\mathbf{w}^* \cdot \mathbf{D} \cdot \mathbf{w} = \frac{1}{2\pi} \int d^2\mathbf{k} \sum_{\substack{n\mathbf{k}\in F \\ m\mathbf{k}\in E}} \frac{\left| \langle u_{m\mathbf{k}} | \hat{\mathbf{v}}_{\mathbf{k}} | u_{n\mathbf{k}} \rangle \cdot \mathbf{w} \right|^2}{\left[ E_{mn,\mathbf{k}} \right]^2} \geq 0 . \tag{4}$$

This confirms that $\mathbf{D} = \left[ d_{ij} \right]_{i,j=1,2}$ is a non-negative tensor. The Chern number in Eq. (2) can be expressed in terms of the elements of $\mathbf{D}$ as follows:

$$\mathcal{C}_F = i\left( d_{12} - d_{21} \right). \tag{5}$$

As $\mathbf{D}$ is non-negative, we have $\begin{pmatrix} 1 & \mp i \end{pmatrix} \cdot \mathbf{D} \cdot \begin{pmatrix} 1 \\ \pm i \end{pmatrix} \geq 0$, which is equivalent to the inequality:

$$d_{11} + d_{12} \pm i\left( d_{12} - d_{21} \right) \geq 0 . \tag{6}$$

Hence, combining the above formula with Eq. (5), we obtain Roy's inequality [14, 22]:

$$\left| \mathcal{C}_F \right| \leq d_{11} + d_{12} . \tag{7}$$

This leads to:

$$\left|\mathcal{C}_F\right| \le \frac{1}{2\pi}\int d^2\mathbf{k} \sum_{\substack{n\mathbf{k}\in F\\ m\mathbf{k}\in E\\ i=1,2}} \left\langle u_{n\mathbf{k}}\left|\hat{v}_{i,\mathbf{k}}\right|u_{m\mathbf{k}}\right\rangle\left\langle u_{m\mathbf{k}}\left|\hat{v}_{i,\mathbf{k}}\right|u_{n\mathbf{k}}\right\rangle \frac{1}{\left[E_{mn,\mathbf{k}}\right]^2}. \quad (8)$$

In this subsection, we suppose that the bands in *F* lie entirely below those in *E*, i.e., *F* contains all bands below the relevant gap, whereas *E* contains all bands above it. We introduce the direct gap width defined as:

$$\Delta_{\mathrm{g}} = \min_{\mathbf{k}}\left|E_{m,\mathbf{k}} - E_{n,\mathbf{k}}\right| \qquad \text{with} \quad n\mathbf{k}\in F, \text{ and } m\mathbf{k}\in E. \quad (9)$$

By construction, $\Delta_{\mathrm{g}}$ gives the minimum separation between the two sets of bands when both are evaluated at the same $\mathbf{k}$.

As $E_{mn,\mathbf{k}} > \Delta_{\mathrm{g}} > 0$, the inequality in Eq. (8) implies that:

$$\left|\mathcal{C}_F\right| \le \frac{1}{\Delta_{\mathrm{g}}}\frac{1}{2\pi}\int d^2\mathbf{k} \sum_{\substack{n\mathbf{k}\in F\\ m\mathbf{k}\in E\\ i=1,2}} \left\langle u_{n\mathbf{k}}\left|\hat{v}_{i,\mathbf{k}}\right|u_{m\mathbf{k}}\right\rangle\left\langle u_{m\mathbf{k}}\left|\hat{v}_{i,\mathbf{k}}\right|u_{n\mathbf{k}}\right\rangle \frac{1}{E_{mn,\mathbf{k}}}. \quad (10)$$

The right-hand side of the above equation can now be evaluated in closed form with the help of the "f-sum rule" [23] (see Appendix A for more details):

$$\frac{1}{2\pi}\int d^2\mathbf{k} \sum_{\substack{n\mathbf{k}\in F\\ m\mathbf{k}\in E}} \left\langle u_{n\mathbf{k}}\left|\hat{v}_{i,\mathbf{k}}\right|u_{m\mathbf{k}}\right\rangle\left\langle u_{m\mathbf{k}}\left|\hat{v}_{i,\mathbf{k}}\right|u_{n\mathbf{k}}\right\rangle \frac{1}{E_{mn,\mathbf{k}}} = \frac{1}{4\pi}\int d^2\mathbf{k} \sum_{n\mathbf{k}\in F} \left\langle u_{n\mathbf{k}}\left|\partial_i^2 \hat{H}_{\mathbf{k}}\right|u_{n\mathbf{k}}\right\rangle, \quad (11)$$

where $\partial_i^2 \equiv \partial / \partial k_i^2$. Let $\left\|...\right\|_\infty$ represent the spectral norm of an operator, i.e., the largest absolute value among its eigenvalues. Then, it follows that

$$\sum_{n\mathbf{k}\in F} \left\langle u_{n\mathbf{k}}\left|\partial_1^2 \hat{H}_{\mathbf{k}} + \partial_2^2 \hat{H}_{\mathbf{k}}\right|u_{n\mathbf{k}}\right\rangle \le N_{\mathrm{F}} \left\|\partial_1^2 \hat{H}_{\mathbf{k}} + \partial_2^2 \hat{H}_{\mathbf{k}}\right\|_\infty, \quad (12)$$

where $N_{\mathrm{F}}$ is the number of bands in set *F*. Combining the above expression with Eqs. (10)-(11), we obtain the following generalization of the Onishi-Fu bound [12]:

$$\left|\mathcal{C}_F\right| \leq \frac{1}{\Delta_{\mathrm{g}}} \frac{\pi}{\mathrm{A}_{\mathrm{cell}}} N_{\mathrm{F}} \max\left\|\nabla_{\mathbf{k}}^2 \hat{H}_{\mathbf{k}}\right\|_{\infty}. \tag{13}$$

In the above, $\max\left\|\nabla_{\mathbf{k}}^2 \hat{H}_{\mathbf{k}}\right\|_{\infty}$ represents the maximum of the "curvature" spectral norm in the Brillouin zone. We took into account that $\int_{BZ} d^2\mathbf{k} = \frac{(2\pi)^2}{\mathrm{A}_{\mathrm{cell}}}$ with $\mathrm{A}_{\mathrm{cell}}$ the area of the direct lattice unit cell. The operator $\nabla_{\mathbf{k}}^2 \hat{H}_{\mathbf{k}}$ may depend explicitly on the unit cell position. In such cases, the maximization of the spectral norm must be done over all the materials in the system.

The derived bound implies that the topological charge is fundamentally limited by the inverse of the direct gap width, the number of bands, and the maximum operator "curvature" per unit of area. In particular, the formula suggests that nontrivial topologies are incompatible with arbitrarily large band gaps [12].

In the context of the Schrodinger equation, the dependence of $\hat{H}_{\mathbf{k}}$ on the wave vector takes the form: $\frac{1}{2m}(\hat{\mathbf{p}} + \hbar\mathbf{k} - q\mathbf{A})^2$. Here, $\hbar$ is the Planck constant, $q$ is the electron charge, $\hat{\mathbf{p}}$ is the momentum operator, $\mathbf{A}$ is the vector potential, and $m$ is the electron mass. Thus, in such a case we have $\max\left\|\nabla_{\mathbf{k}}^2 \hat{H}_{\mathbf{k}}\right\|_{\infty} = \frac{2\hbar^2}{m}$, i.e., the curvature is determined mainly by the electron mass. Then, the bound in Eq. (13) reduces to the Onishi-Fu result: $\left|\mathcal{C}_F\right| \leq \frac{1}{\Delta_{\mathrm{g}}} \frac{2\pi\hbar^2}{m} n_{\mathrm{F}}$, with $n_{\mathrm{F}} = \frac{N_{\mathrm{F}}}{\mathrm{A}_{\mathrm{cell}}}$ the electron density [12].

It is interesting to note that if $\hat{H}_{\mathbf{k}}$ is represented by a matrix (e.g., the Haldane model) then the total number of bands is finite. This means that the spectrum of $-\hat{H}_{\mathbf{k}}$ also has a lower-bound. Hence, for a given band gap, the operation $\hat{H}_{\mathbf{k}} \to -\hat{H}_{\mathbf{k}}$ interchanges the filled and empty bands, and thus transforms the gap Chern number as $\mathcal{C} \to -\mathcal{C}$ . Thus, we have the additional bound:

$$\left|\mathcal{C}_F\right| \le \frac{1}{\Delta_{\mathrm{g}}} \frac{\pi}{\mathrm{A}_{\mathrm{cell}}} N_{\mathrm{E}} \max \left\| \nabla_{\mathbf{k}}^2 \hat{H}_{\mathbf{k}} \right\|_{\infty}, \tag{14}$$

where $N_{\mathrm{E}}$ stands for the number of bands above the gap.

*B. Velocity bound*

Even though the curvature bound (13) discussed in the previous section is rather powerful, it has some limitations. The main one is that it depends explicitly on the number of bands below the gap, $N_{\mathrm{F}}$. In some bosonic systems, however, this number is infinite [24, 25], which renders the bound ineffective. This occurs because typical bosonic fields satisfy a reality symmetry, for example, the electromagnetic field is real valued. As a result, the spectrum of a photonic crystal contains both positive- and negative-frequency branches and, unlike in electronic systems, there is no effective ground state [24, 25]. Another limitation worth mentioning is that the bound is proportional to the size of the Brillouin zone ($(2\pi)^2 / \mathrm{A}_{\mathrm{cell}}$). Consequently, it is not applicable to continuum systems with no intrinsic periodicity [26, 27].

Next, we derive a simple bound that overcomes the first deficiency discussed above. This alternative bound is still expressed in terms of the number of bands in the set *F*, but, unlike the "curvature" bound of the previous subsection, it does not require *F* to include all the bands below the gap. Hence, this bound can be useful, for example, to estimate the maximum topological charge of an arbitrary finite set of bands in a bosonic system with reality symmetry.

Our starting point is to rewrite Eq. (2), in the following way:

$$\mathcal{C}_F = \frac{i}{2\pi} \sum_{n\mathbf{k}\in F} \int d^2\mathbf{k} \left\langle u_{n\mathbf{k}} \right| \mathbf{A}_{n\mathbf{k}}^{12} \left| u_{n\mathbf{k}} \right\rangle - \left\langle u_{n\mathbf{k}} \right| \mathbf{A}_{n\mathbf{k}}^{21} \left| u_{n\mathbf{k}} \right\rangle . \tag{15}$$

where $\mathbf{A}_{n\mathbf{k}}^{ij} = \hat{v}_{\mathbf{k},i} G_{\mathbf{k},n} \hat{v}_{\mathbf{k},j}$ with $G_{\mathbf{k},n} = \sum_{m\mathbf{k}\in E} \frac{1}{\left[E_{mn,\mathbf{k}}\right]^2} \left| u_{m\mathbf{k}} \right\rangle \left\langle u_{m\mathbf{k}} \right|$.

Taking into account that $\left| \left\langle u_{n\mathbf{k}} \right| \mathbf{A}_{n\mathbf{k}}^{ij} \left| u_{n\mathbf{k}} \right\rangle \right| \le \left\| \mathbf{A}_{n\mathbf{k}}^{ij} \right\|_\infty \left\| u_{n\mathbf{k}} \right\|^2$ and that the spectral norm satisfies $\left\| \mathbf{A}_1 \mathbf{A}_2 .... \mathbf{A}_N \right\|_\infty \le \left\| \mathbf{A}_1 \right\|_\infty ... \left\| \mathbf{A}_N \right\|_\infty$ [28], we find that:

$$\left| \mathcal{C}_F \right| \le 2 \times \frac{1}{2\pi} \sum_{n\mathbf{k}\in F} \int d^2\mathbf{k} \left\| \hat{v}_{\mathbf{k},1} \right\|_\infty \left\| \hat{v}_{\mathbf{k},2} \right\|_\infty \left\| G_{\mathbf{k},n} \right\|_\infty , \tag{16}$$

where we used $\left\| u_{n\mathbf{k}} \right\| = 1$. Clearly, all the eigenvalues of $G_{\mathbf{k},n}$ are bounded by $\frac{1}{\Delta_{\mathrm{g}}^2}$ with $\Delta_{\mathrm{g}}$ the direct gap width defined as in Eq. (9). This observation leads to:

$$\left| \mathcal{C}_F \right| \le \frac{1}{\Delta_{\mathrm{g}}^2} \frac{4\pi N_{\mathrm{F}}}{\mathrm{A}_{\mathrm{cell}}} \max \left\| \partial_1 \hat{H}_{\mathbf{k}} \right\|_\infty \left\| \partial_2 \hat{H}_{\mathbf{k}} \right\|_\infty , \tag{17}$$

where as before $N_{\mathrm{F}}$ is the number of bands in set *F*. Different from the previous subsection, the present derivation does not require *F* to be formed by the lowest frequency bands: it can be any subset of bands. Thus, typically $\Delta_{\mathrm{g}}$ can be regarded as the minimum of the two independent direct gaps: $\Delta_{\mathrm{g}} = \min_{\mathbf{k}} \left\{ \Delta_{\mathrm{gU},\mathbf{k}}, \Delta_{\mathrm{gL},\mathbf{k}} \right\}$, where $\Delta_{\mathrm{gU},\mathbf{k}}$ is the gap between the set *F* and all the bands above it (if any) at a fixed $\mathbf{k}$, whereas $\Delta_{\mathrm{gL},\mathbf{k}}$ is the gap between the set *F* and all the bands below it (if any) evaluated at the same $\mathbf{k}$.

We note in passing that the bound (17) could be slightly refined, specifically one could replace $\frac{N_{\mathrm{F}}}{\Delta_{\mathrm{g}}^2}$ by $\sum_{n\in F} \frac{1}{\Delta_{\mathrm{g},n}^2}$ where $\Delta_{\mathrm{g},n} = \min_{\mathbf{k}} \left| E_{m,\mathbf{k}} - E_{n,\mathbf{k}} \right|$ with $m\mathbf{k} \in E$ is the minimum "distance" between band *n* and the bands in set *E*.

We shall refer to Eq. (17) as the "velocity bound", as it constrains the maximum topological charge to the spectral norm of velocity operators. This bound is especially relevant in the bosonic case, where the operator $\hat{H}_{\mathbf{k}}$ is often linear in the momentum. For example, Maxwell theory is intrinsically relativistic, and, as a consequence, the corresponding velocity operators $\hat{v}_{\mathbf{k},i} = \partial_i \hat{H}_{\mathbf{k}}$ are typically independent of the wave vector and bounded by the speed of light: $\left\|\hat{v}_{\mathbf{k},i}\right\|_\infty < c$. Therefore, the velocity bound (17) provides a natural constraint for the topological charge of systems whose velocity is fundamentally limited, as in relativistic theories. For example, for realistic electromagnetic dispersive systems, we have (it is implicit that $\hat{H}_{\mathbf{k}}$ has units of frequency):

$$\left|\mathcal{C}_F\right| \le \frac{1}{\Delta_{\mathrm{g}}^2} \frac{4\pi N_{\mathrm{F}}}{\mathrm{A}_{\mathrm{cell}}} c^2 . \tag{18}$$

It is relevant to note that condensed matter systems are described by the Schrödinger equation, which being nonrelativistic does not impose a natural bound on the velocity operator.

As shown in Appendix B, the velocity bound can be extended to non-Hermitian systems [29, 30, 31]. In particular, in the non-Hermitian case the right-hand side of Eq. (17) must be multiplied by an additional factor, $\kappa_{\max}^2$, given by the maximum condition number ($\kappa_{\max} = \max_{\mathbf{k}} \kappa_{\mathbf{k}}$ with $\kappa_{\mathbf{k}} = \left\|R_{\mathbf{k}}\right\|_\infty \left\|R_{\mathbf{k}}^{-1}\right\|_\infty$) of the eigenvector matrix (see Eq. (B8)). Here, the eigenvector matrix $R_{\mathbf{k}}$ is the matrix (or operator) whose columns are the eigenvectors of the Hamiltonian. This condition number always satisfies a lower bound of 1 ($\kappa_{\mathbf{k}} \ge 1$), and for Hermitian operators it is exactly 1. By contrast, near an exceptional point it can become very large. Although evaluating this quantity is often impractical, this result suggests that, in non-Hermitian systems, the constraints on the topological charge are weaker than in the Hermitian case.

## III. APPLICATION OF THE CURVATURE BOUND

Next, we validate the theoretical formalism through numerical examples. Specifically, in this section we apply the curvature bound (13) to constraint the topological charge of a few photonic crystals.

### *A. Ferrite photonic crystal*

As a first example, we consider a magnetically gyrotropic photonic crystal formed by cylindrical air rods embedded in a magnetized ferrite background [Fig. 1a]. The Maxwell's equations in this system read as:

$$\nabla \times \mathbf{E} = i\omega\mu_0 \overline{\mu} \cdot \mathbf{H}, \qquad \nabla \times \mathbf{H} = -i\omega\varepsilon_0 \varepsilon \mathbf{E}, \tag{19}$$

where the relative permittivity and permeability tensors take the form:

$$\overline{\varepsilon} = \varepsilon \mathbf{1}_{3\times3}, \qquad \overline{\mu} = \begin{pmatrix} \mu & i\kappa & 0 \\ -i\kappa & \mu & 0 \\ 0 & 0 & 1 \end{pmatrix}, \tag{20}$$

being $\kappa$ the nonreciprocity parameter. This type of material response occurs in natural ferrimagnetic materials (e.g., ferrites) biased with a magnetic field directed along the $z$-direction. The material dispersion is neglected in this subsection. This is done by replacing the frequency-dependent material parameters with their values at a fixed frequency $\omega_0$, where $\omega_0$ is chosen in the spectral range of interest.

We consider waves with transverse electric polarization ($\mathbf{E} = E_z \hat{\mathbf{z}}$ and $\mathbf{H} = H_x \hat{\mathbf{x}} + H_y \hat{\mathbf{y}}$) and propagation in the *x*-*y* plane, $E_z = E_z(x, y)$. In such a case, the Maxwell's equations can be reduced to a standard scalar spectral problem $\hat{H} \cdot \Psi = \left(\frac{\omega}{c}\right)^2 \Psi$, with $\Psi = \varepsilon^{1/2} E_z$ and $\hat{H}$ an Hermitian operator such that [31]:

$$\hat{H} = -\varepsilon^{-1/2} \partial_x \left( \mu_{\text{ef}}^{-1} \partial_x - i \chi \partial_y \right) \varepsilon^{-1/2} - \varepsilon^{-1/2} \partial_y \left( \mu_{\text{ef}}^{-1} \partial_y + i \chi \partial_x \right) \varepsilon^{-1/2}, \tag{21}$$

with $\mu_{\text{ef}} = \left(\mu^2 - \kappa^2\right) / \mu$ and $\chi = \kappa / \left(\mu^2 - \kappa^2\right)$. It is implicit that the material parameters can be position dependent, e.g., $\varepsilon = \varepsilon(x, y)$.

As usual, the Bloch modes with wave vector $\mathbf{k} = k_x \hat{\mathbf{x}} + k_y \hat{\mathbf{y}}$ are determined by a family of operators $\hat{H}_{\mathbf{k}}$, which can be obtained from $\hat{H}$ with the replacements $\partial_x \to \partial_x + ik_x$ and $\partial_y \to \partial_y + ik_y$. It is clear from Eq. (21) that $\hat{H}_{\mathbf{k}}$ exhibits a quadratic dependence on the wave vector $\mathbf{k}$. In particular, it is readily found that the Laplacian of the operator with respect to $\mathbf{k}$ is given by $\nabla_{\mathbf{k}}^2 \hat{H}_{\mathbf{k}} = \frac{4}{\varepsilon \mu_{\text{ef}}}$. Thus, the curvature bound (13) applied to this system reduces to:

$$\left| \mathcal{C}_F \right| \le C_{\text{B}}, \quad \text{with} \quad C_{\text{B}} = \frac{1}{\Delta_{\text{g}}} \frac{4\pi}{\text{A}_{\text{cell}}} N_{\text{F}} \max \frac{1}{\varepsilon \mu_{\text{ef}}}. \tag{22}$$

where the maximum is evaluated over the unit cell spatial domain.

In nondispersive crystals, the eigenvalues of $\hat{H}_{\mathbf{k}}$ are typically related to $\left(\frac{\omega}{c}\right)^2$, as in the present example, rather than directly to the frequency. Due to this reason, the direct gap width is given by, $\Delta_{\mathrm{g}} = \min\left(\frac{\omega_{E,\mathbf{k}}^2}{c^2} - \frac{\omega_{F,\mathbf{k}}^2}{c^2}\right)$, with the indices "E" and "F" referring to the "empty" and "filled" bands that delimit the gap.

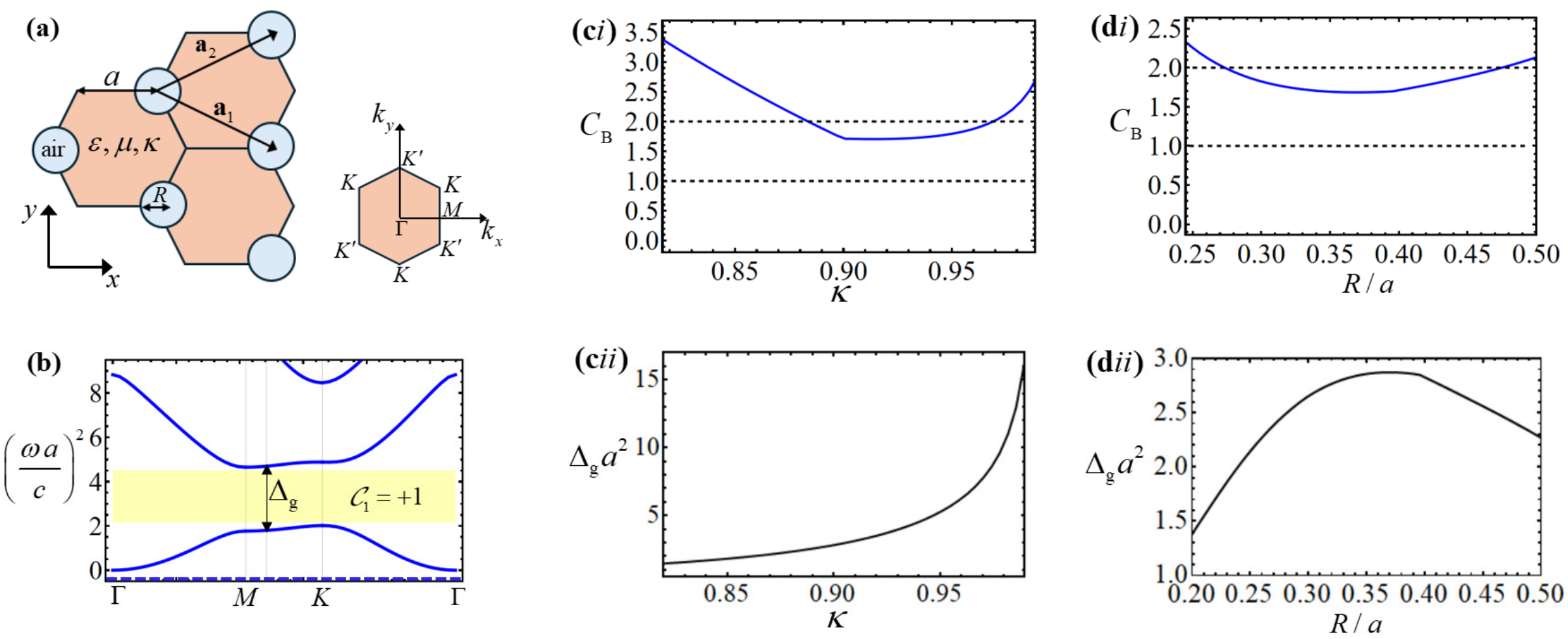

**Fig. 1** **(a)** Geometry of the ferrite photonic crystal formed by cylindrical air rods with radius $R$ embedded in a nondispersive ferrite background characterized by with parameters $\varepsilon$, $\mu$ and $\kappa$. The inset shows the Brillouin zone and the high symmetry points $\Gamma, M$ and $K$. The direct lattice primitive vectors are $\mathbf{a}_1 = a/2\left(3,\sqrt{3}\right)$ and $\mathbf{a}_2 = a/2\left(3,-\sqrt{3}\right)$, with $a$ the length represented on the figure. **(b)** Band structure of the nondispersive ferrite photonic crystal. The air scatterers have a radius of $R = 0.4a$. The material parameters of the ferrite background are taken as $\mu = 1$, $\kappa = 0.9$ and $\varepsilon = 1/\mu_{\mathrm{ef}} = 5.26$. The band gap indicated by the yellow horizontal shaded strip has a topological charge of $\mathcal{C}_1 = +1$. The direct gap width is $\Delta_{\mathrm{g}}$. The dashed blue horizontal line in the dispersion plot represents the flat degenerate bands at zero frequency, which arise in the vector formulation of the problem. **(ci)** Bound on the gap Chern number, $C_{\mathrm{B}}$, calculated using Eq. (22) for varying $\kappa$. The remaining material parameters are the same as in panel (b). The dashed horizontal gray lines mark transitions in the integer part of $C_{\mathrm{B}}$ and are included for visual guidance. **(cii)** Direct gap width $\Delta_{\mathrm{g}}$ for varying $\kappa$. **(d)** Similar to (c) but for varying radius $R/a$.

To illustrate the application of the bound (22), we consider a ferrite photonic crystal formed by air rods with radius $R = 0.4a$ organized in a hexagonal lattice in a nondispersive ferrite background characterized by with parameters $\mu = 1$, $\kappa = 0.9$ and $\varepsilon = 1/\mu_{\rm ef} = 5.26$, as sketched in Fig. 1a. The band structure of the photonic crystal and the respective band gap are represented in Fig. 1b. The band structure is computed numerically using the plane wave method. Here and in the rest of the examples of the article, we used 49 plane waves to expand each scalar component of the state vector. Topological analysis based on a Green's function method developed in previous work [31, 32], reveals that the topological charge associated with the relevant band gap is $\mathcal{C}_1 = +1$.

It is well established that the gap Chern number is invariant under continuous deformations of the photonic crystal geometry, as long as the corresponding band gap remains open. In contrast, the upper bound $C_{\rm B}$ in Eq. (22) is geometry dependent. We take advantage of this property to derive a tight bound for the gap Chern number of the photonic crystal.

Figure 1c shows the bound $C_{\rm B}$ [Fig. 1ci] and the corresponding direct gap width $\Delta_{\rm g}$ [Fig. 1cii], computed using Eq. (22) with $N_{\rm F} = 1$ and $A_{\rm cell} = |\mathbf{a}_1 \times \mathbf{a}_2| = \frac{3\sqrt{3}}{2} a^2$, as a function of the ferrite nonreciprocal parameter $\kappa$. The remaining parameters are the same as those used in Fig. 1b. Note that the curve representing $C_{\rm B}$ can exhibit kink-like features because the wave vector associated with the minimum direct gap can change discontinuously as the system parameters vary. As seen in Fig. 1ci, the tightest bound is achieved for $\kappa = 0.9$ and is approximately 1.7, which constrains the gap Chern number of the original photonic crystal to the values 0, +1, or −1. Since the exact Chern number is $\mathcal{C}_1 = +1$, the upper bound is sufficiently tight to effectively constrain the topological charge using only the crystal band structure, i.e., without performing explicitly any topological calculation.

Even though our crystal is topological, Fig. 1cii shows that the gap width can be fairly wide for large $\kappa$, seemingly contradicting Eq. (22). The resolution is that the term $\max\frac{1}{\varepsilon\mu_{\text{ef}}}=\max\frac{\mu}{\varepsilon\left(\mu^2-\kappa^2\right)}$ in Eq. (22) becomes also large when $\kappa\to\mu$, allowing the crystal to remain topological despite the large gap width. Moreover, the bound can be further optimized by varying other system parameters, such as the radius of the rods, as illustrated in the plots of Fig. 1d.

It is interesting to note that, by choosing the magnetic field as the state vector rather than the electric field, the same propagation problem can be formulated using a different $\hat{H}$. Indeed, from the Maxwell's equations (19) one can write $\nabla\times\frac{1}{\varepsilon}\nabla\times\mathbf{H}=\frac{\omega^2}{c^2}\bar{\mu}\cdot\mathbf{H}$. Assuming that $\bar{\mu}$ is positive definite (i.e., that $\mu>|\kappa|$), this can be expressed as $\hat{H}\cdot\Psi=\frac{\omega^2}{c^2}\Psi$ where $\Psi=\bar{\mu}^{1/2}\cdot\mathbf{H}$ and $\hat{H}$ is the Hermitian operator:

$$\hat{H}=\bar{\mu}^{-1/2}\cdot\left(\nabla\times\frac{1}{\varepsilon}\nabla\times\mathbf{1}\right)\cdot\bar{\mu}^{-1/2}. \tag{23}$$

For 2D problems, $\hat{H}$ only acts on the first two (*x* and *y*) components of the magnetic field. As before, the Bloch operator $\hat{H}_{\mathbf{k}}$ is obtained from $\hat{H}$ with the replacements $\partial_x\to\partial_x+ik_x$ and $\partial_y\to\partial_y+ik_y$. Straightforward calculations show that in this case $\nabla_{\mathbf{k}}^2\hat{H}_{\mathbf{k}}=\frac{2}{\varepsilon}\bar{\mu}^{-1}$. Substituting in Eq. (13) and using $\left\|\bar{\mu}^{-1}\right\|_\infty=\frac{1}{\mu-|\kappa|}$, we get the alternative bound:

$$\left|\mathcal{C}_F\right| \le C_{\mathrm{B}}\text{, with}\quad C_{\mathrm{B}} = \frac{1}{\Delta_{\mathrm{g}}}\frac{2\pi}{\mathrm{A}_{\mathrm{cell}}}N_{\mathrm{F}}\max\frac{1}{\varepsilon}\frac{1}{\mu-|\kappa|}. \tag{24}$$

Interestingly, although the two propagation problems are formally equivalent, the bounds obtained for the two operators are different, as demonstrated in the corresponding formulas, Eq. (22) and Eq. (24). Moreover, one can readily verify that for $\mu > |\kappa|$ , one has $\max\frac{1}{\varepsilon}\frac{1}{\mu-|\kappa|} \le \max\frac{2}{\varepsilon}\frac{1}{\mu_{\mathrm{ef}}}$ , suggesting that the bound provided by (24), i.e., the vector formulation, is tighter than the bound (22) provided by the scalar formulation.

There is, however, an important subtlety: even though the two operators describe the same propagation problem, their spectra do not coincide exactly. In addition to the Bloch modes described by (21), the operator in Eq. (23) also supports a continuum of longitudinal-type modes at zero frequency $\omega = 0$. Specifically, these modes are associated with solutions of the form $\Psi = \bar{\mu}^{1/2}\cdot\nabla\phi$ (or equivalently $\mathbf{H} = \nabla\phi$), where $\phi$ is an arbitrary function of position. Due to the mechanisms of band folding, the longitudinal-type modes result in an *infinite* number of degenerate zero-frequency bands (see the dashed blue line in the band structure of Fig. 1b).

This property creates a difficulty: in the vector problem, the number of bands below the relevant gap is no longer equal to one, in fact, it is not even finite. Although the longitudinal bands are not topologically relevant, they cannot be ignored in the estimation of the curvature bound. In particular, the f-sum rule requires all bands below the gap to be included. Therefore, the presence of infinitely many bands below the gap, $N_{\mathrm{F}} = \infty$, renders the bound in Eq. (24) trivial.

In summary, the previous analysis illustrates the importance of having only a finite number of bands below the gap. It also shows how longitudinal waves in bosonic systems can violate this requirement, highlighting the need to formulate the physical problem in a way that avoids such spurious modes.

### *B. Photonic crystal with a magnetoelectric Tellegen-type response*

In a second example, we consider a photonic crystal composed of air rods embedded in a metallic background exhibiting a nondispersive Tellegen-type response [Fig. 2a]. The photonic crystal is characterized by a position dependent response of the type [33, 34]:

$$\begin{pmatrix} \mathbf{D} \\ \mathbf{B} \end{pmatrix} = \mathbf{M} \cdot \begin{pmatrix} \mathbf{E} \\ \mathbf{H} \end{pmatrix}, \quad \text{with} \quad \mathbf{M} = \begin{pmatrix} \varepsilon_0 \overline{\varepsilon} & \frac{1}{c} \overline{\xi} \\ \frac{1}{c} \overline{\zeta} & \mu_0 \overline{\mu} \end{pmatrix}, \tag{25}$$

where the relative permittivity, permeability, and magnetoelectric coupling tensors are denoted by $\overline{\varepsilon}$, $\overline{\mu}$, $\overline{\xi}$, $\overline{\zeta}$, respectively. The permittivity in the background region is described by a Drude-type model $\varepsilon = 1 - \frac{\omega_{\mathrm{p}}^2}{\omega^2}$, and in the air rods by $\varepsilon = 1$. Here, $\omega_{\mathrm{p}}$ is the plasma frequency. Furthermore, the permeability response is assumed trivial, $\overline{\mu} = \mathbf{1}_{3\times3}$. Finally, the magnetoelectric tensors are taken as $\overline{\zeta} = \overline{\xi} = \hat{\mathbf{z}} \otimes \boldsymbol{\xi} + \boldsymbol{\xi} \otimes \hat{\mathbf{z}}$, with $\boldsymbol{\xi} = \boldsymbol{\xi}(\mathbf{r})$ a vector that plays the role of a magnetic potential for photons [33, 34]. Similar to Refs. [33, 34], we assume that:

$$\boldsymbol{\xi}(\mathbf{r}) = \xi_0 \frac{\sqrt{3}a}{4\pi} \left[ \mathbf{b}_1 \sin(\mathbf{b}_1 \cdot \mathbf{R}) + \mathbf{b}_2 \sin(\mathbf{b}_2 \cdot \mathbf{R}) + (\mathbf{b}_1 + \mathbf{b}_2) \sin([\mathbf{b}_1 + \mathbf{b}_2] \cdot \mathbf{R}) \right]. \tag{26}$$

Here, $\xi_0$ denotes the (dimensionless) amplitude of the Tellegen-type vector which determines the strength of the nonreciprocal coupling. In the above, $\mathbf{b}_1 = \frac{2\pi}{a}\left(\frac{1}{3}, -\frac{1}{\sqrt{3}}\right)$ and $\mathbf{b}_2 = \frac{2\pi}{a}\left(\frac{1}{3}, \frac{1}{\sqrt{3}}\right)$ denote the reciprocal lattice primitive vectors, $\mathbf{R} = \mathbf{r} - \mathbf{r}_c$ where $\mathbf{r}_c = 1/3(\mathbf{a}_1 + \mathbf{a}_2)$ determines the coordinates of the honeycomb cell's center, and $\mathbf{a}_1 = a/2\left(3, \sqrt{3}\right)$ and $\mathbf{a}_2 = a/2\left(3, -\sqrt{3}\right)$ are the direct lattice primitive vectors [Fig. 2a]. In a previous study, we have demonstrated that this system provides a photonic analogue of the Haldane model [33, 34]. For more details about this Tellegen crystal, a reader is referred to these works.

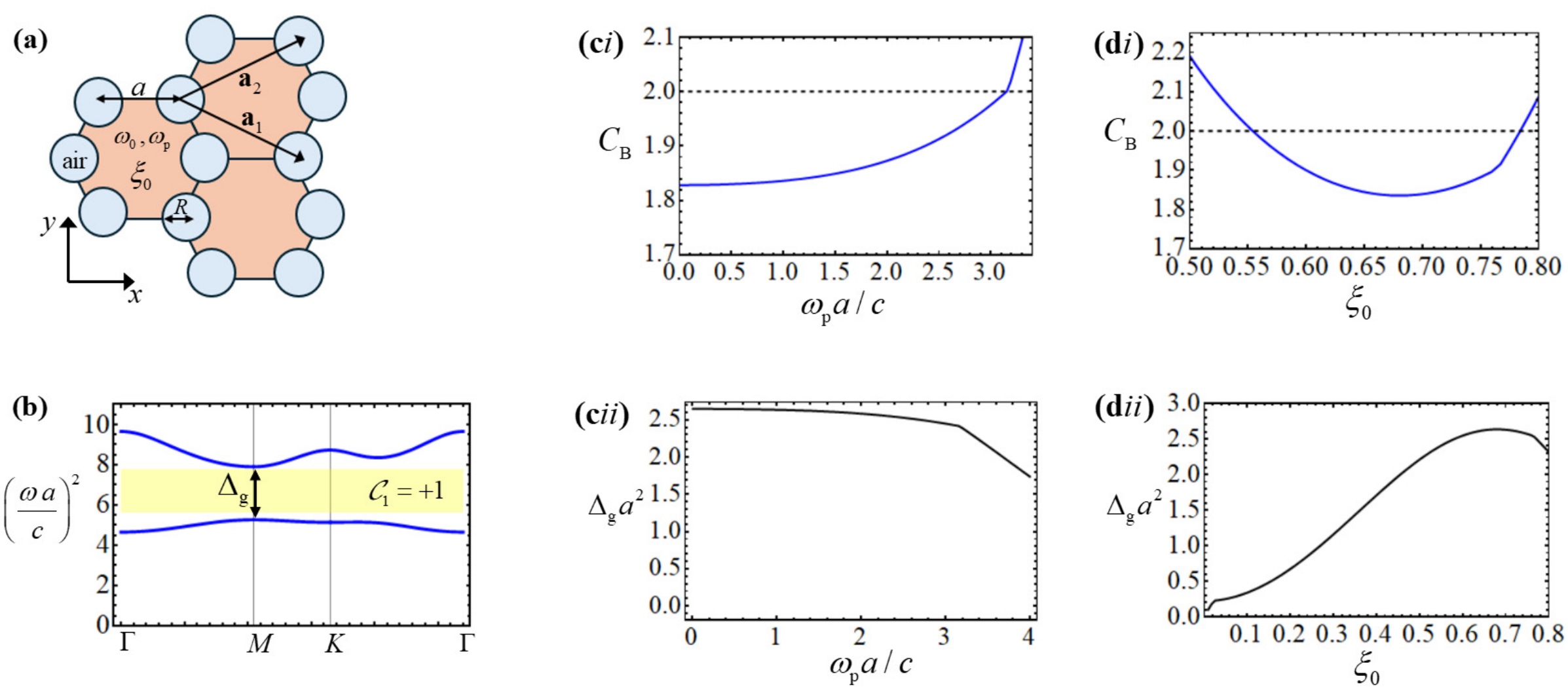

**Fig. 2 (a)** Geometry of a photonic crystal formed by cylindrical air rods with radius $R$ embedded in a metallic background, characterized by the plasma frequency $\omega_p$. The overall strength the position-dependent Tellegen coupling applied to the crystal is determined by the parameter $\xi_0$. **(b)** Band structure of the Tellegen-type photonic crystal, for the material parameters $\omega_p a/c = 1$, $\xi_0 = 0.67$, and $\omega_0 a/c = 4.38$. The radius of the scatterers is $R = 0.35a$. **(ci)** Bound on the gap Chern number calculated using Eq. (28), for a varying plasma frequency $\omega_p$. The remaining material parameters are the same as in panel (b). The dashed horizontal gray line marks the transition in the integer part of $C_B$. **(cii)** Corresponding direct gap width $\Delta_g$ for a varying plasma frequency $\omega_p$. **(d)** Analogous to (c), but for a varying Tellegen parameter $\xi_0$.

Similar to the previous subsection, we restrict our attention to transverse electric waves with $\mathbf{E} = E_z \hat{\mathbf{z}}$ and $\mathbf{H} = H_x \hat{\mathbf{x}} + H_y \hat{\mathbf{y}}$, and obtain a wave equation for the electric field. A straightforward analysis, yields $\left[-\left(\nabla - i\frac{\omega}{c}\hat{\mathbf{z}}\times\boldsymbol{\xi}(\mathbf{r})\right)^2 + \frac{\omega_{\rm p}^2(\mathbf{r})}{c^2}\right]E_z = \frac{\omega^2}{c^2}E_z$ [33, 34]. As can be seen, the operator on the left-hand side is frequency dependent, which prevents a strict formulation of the Bloch problem as a standard scalar eigenvalue problem. This obstruction can be attributed in part to the dispersive nature of the Drude background, and in part to the magneto-electric response associated with the Tellegen coupling.

In order to circumvent this difficulty, in this subsection we adopt a simple remedy: we replace the frequency $\omega$ appearing in the frequency-dependent scalar operator by a constant $\omega_0$. This approximation transforms the wave equation into a standard eigenvalue problem ($\hat{H}\cdot E_z = \frac{\omega^2}{c^2}E_z$) and avoids complications arising from material dispersion [33]:

$$\hat{H} = -\left(\nabla - i\frac{\omega_0}{c}\hat{\mathbf{z}}\times\boldsymbol{\xi}(\mathbf{r})\right)^2 + \frac{\omega_{\rm p}^2(\mathbf{r})}{c^2}. \tag{27}$$

Provided that the constant frequency $\omega_0$ lies within the relevant gap of the unperturbed operator, the replacement does not alter the underlying physics, as we shall demonstrate in detail in section IV.

As seen in Eq. (27), the operator is quadratic in the spatial derivatives, which allows us to directly apply the curvature bound to this system. In fact, the Laplacian of $\hat{H}_{\mathbf{k}}$, obtained from $\hat{H}$ with the replacement $\nabla \to \nabla + i\mathbf{k}$, is given by $\nabla_{\mathbf{k}}^2\hat{H}_{\mathbf{k}} = 4$. Thereby, the curvature bound (13) reduces to:

$$|\mathcal{C}_F| \le C_{\rm B}, \text{ with } \quad C_{\rm B} = \frac{1}{\Delta_{\rm g}}\frac{4\pi}{{\rm A}_{\rm cell}}N_{\rm F}. \tag{28}$$

As in the previous subsection, the direct gap width is $\Delta_{\mathrm{g}} = \min\left( \frac{\omega_{E,\mathbf{k}}^2}{c^2} - \frac{\omega_{F,\mathbf{k}}^2}{c^2} \right)$.

To illustrate the application of the bound, we consider Tellegen-type photonic crystal characterized by the parameters $\omega_{\mathrm{p}} a / c = 1$, $\xi_0 = 0.67$, and $\omega_0 a / c = 4.38$. The air rods are organized in a honeycomb lattice with lattice constant *a*. Their radius is $R = 0.35a$. The photonic band structure of this system is plotted in Fig. 2b. A topological characterization shows that the spectrum exhibits a low-frequency topologically nontrivial band gap between the first and second bands with a gap Chern number of $\mathcal{C}_1 = +1$ [33].

Similar to the previous subsection, we show in Figs. 2c and 2d how the bound $C_{\mathrm{B}}$ in Eq. (28) varies with the system parameters, for $N_{\mathrm{F}} = 1$ and $\mathrm{A}_{\mathrm{cell}} = \left| \mathbf{a}_1 \times \mathbf{a}_2 \right| = \frac{3\sqrt{3}}{2} a^2$. As seen, by optimizing the geometry in terms of the parameters $\omega_{\mathrm{p}} a / c$ [Fig. 2ci] and $\xi_0$ [Fig. 2di], it is possible to obtain a tight upper bound, such that the integer part of $C_{\mathrm{B}}$ coincides with the actual gap Chern number of the system.

Furthermore, as discussed next, the bound can capture the evolution of the topological charge across a topological transition in a high-frequency band gap between the second and third bands [Fig. 3a]. Figures 3ai-aiii display the band structure for progressively increasing values of the plasma frequency $\omega_{\mathrm{p}}$. As seen, for $\omega_{\mathrm{p}} a / c = 4$ [Fig. 3aii], the band gap closes, signaling a topological phase transition. Indeed, we numerically checked that for relatively small values of $\omega_{\mathrm{p}} a / c$ [Fig. 3ai], the gap Chern number of the 2$^{\mathrm{nd}}$ gap is identical to that of the 1$^{\mathrm{st}}$ gap, $\mathcal{C}_2 = +1$, whereas for large values of $\omega_{\mathrm{p}} a / c$ [Fig. 3aiii] the gap becomes topologically trivial [$\mathcal{C}_2 = 0$]. As seen in Fig. 3aii, the transition occurs due to exchange of topological charge between the 3$^{\mathrm{rd}}$ band and the lower frequency bands. Note that the fact that $\mathcal{C}_2 = \mathcal{C}_1$ for small values of $\omega_{\rm p} a / c$ implies that the 2nd band has vanishing Chern number.

The corresponding upper bound $C_{\rm B}$ is shown in Fig. 3b as a function of the plasma frequency $\omega_{\rm p} a / c$. As seen, before the phase transition, the bound reaches values below +2, making it sufficiently tight so that its integer part matches the exact topological index of $\mathcal{C}_2 = +1$. For larger parameter values of $\omega_{\rm p} a / c$, the bound drops below +1, indicating that the gap becomes topologically trivial ($\mathcal{C}_2 = 0$) consistent with our full-wave topological analysis.

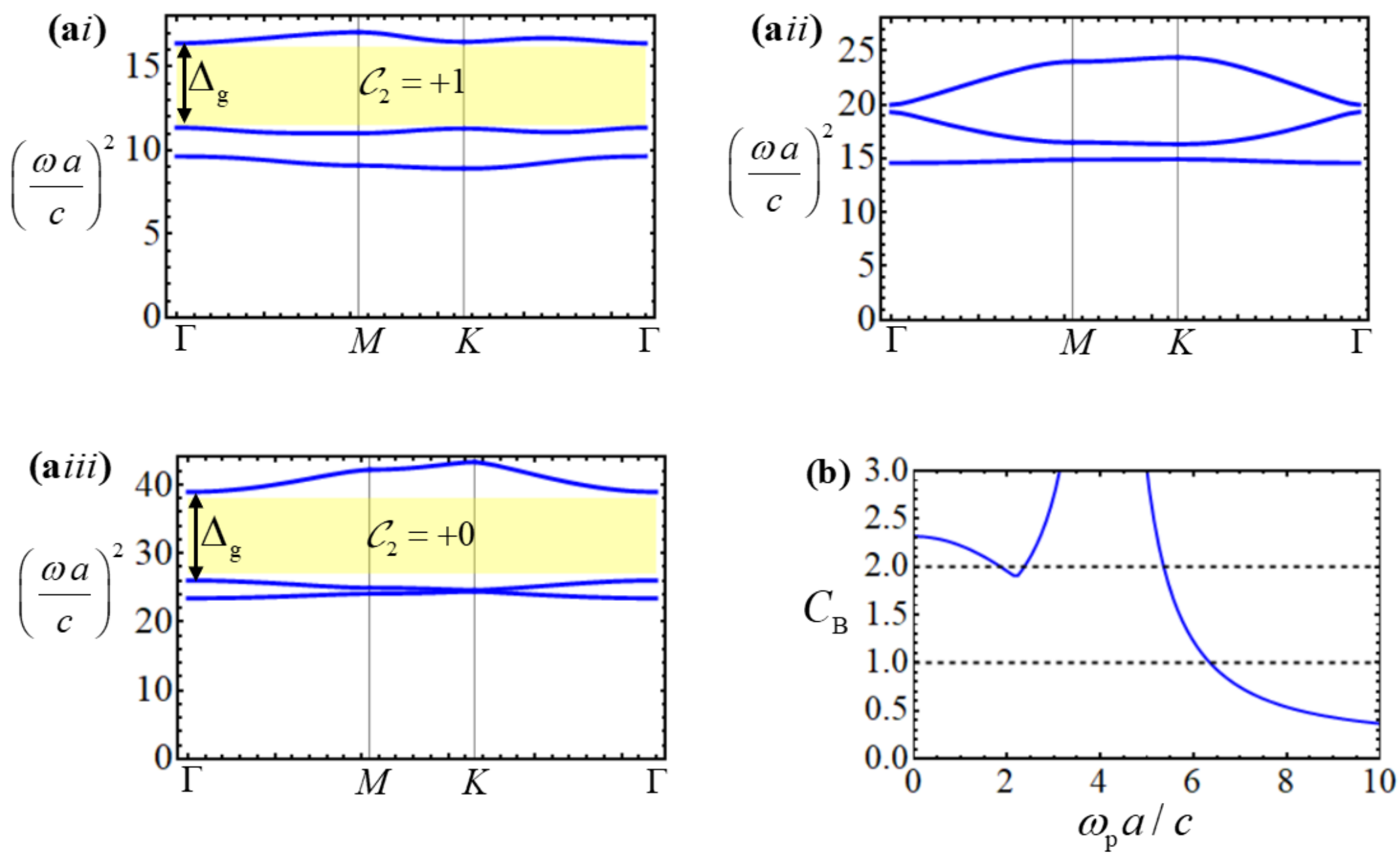

**Fig. 3 (a)** (i)-(iii) Evolution of the band structure of the Tellegen-type photonic crystal for varying $\omega_{\rm p} a / c$. The radius of the scatterers is $R = 0.35a$. The material parameters are $\xi_0 = 0.9$, $\omega_0 a / c = 4.38$ and **(a*i*)** $\omega_{\rm p} a / c = 2.25$, **(a*ii*)** $\omega_{\rm p} a / c = 4$, **(a*iii*)** $\omega_{\rm p} a / c = 7$. The band gap highlighted by with the yellow horizontal shaded strip between the third and second bands has topological charge of $\mathcal{C}_2$. **(b)** Bound on the 2nd gap Chern number as function of the plasma frequency $\omega_{\rm p} a / c$.

## IV. APPLICATION OF THE VELOCITY BOUND

### *A. Dispersive nonreciprocal crystals*

The examples of Sect. III have in common that the wave propagation problem can be reduced to a generalized scalar eigenvalue problem of the type $\hat{H}(\mathbf{r},\nabla,\omega)\cdot\Psi=\left(\frac{\omega}{c}\right)^2\Psi$, where the operator $\hat{H}$ depends explicitly on the frequency. This frequency dependence was dropped in our analysis by fixing $\omega$ ( $\hat{H}(\omega)\to\hat{H}(\omega_0)$ ). Roughly speaking, this procedure removes the frequency dependence of the material parameters, as in the ferrite crystal, or more generally the frequency dependence of the operator itself, as in the Tellegen-crystal example. This allowed us to recover a standard eigenvalue problem of the form $\hat{H}(\mathbf{r},\nabla,\omega_0)\cdot\Psi=\left(\frac{\omega}{c}\right)^2\Psi$, with a spectrum bounded from below, and with a Bloch-operator family $\hat{H}_{\mathbf{k}}=\hat{H}(\mathbf{r},\nabla+i\mathbf{k},\omega_0)$ that is quadratic in the wave vector, as required for the application of the curvature bound. In general, applying the curvature bound to a nonreciprocal photonic crystal inevitably entails a similar procedure.

In fact, the exact spectrum of dispersive (bosonic) photonic crystals is formed by both positive and negative frequency branches due to the reality symmetry of the fields (particle-hole symmetry). [24]. An exact description of the propagation problem necessarily requires an operator whose eigenvalues are the frequencies themselves, rather than their squares. This creates a difficulty: since the spectrum contains both positive and negative frequencies, it is necessarily unbounded from below. Consequently, there are infinitely many bands below any given gap, $N_{\mathrm{F}}=\infty$, which makes the curvature bound in Eq. (13) trivial. This situation is analogous to the one already encountered in Sect. III.A, where the vector formulation led to an infinite number of zero-frequency longitudinal-type flat bands.

To illustrate the discussion, we consider again the Tellegen-type photonic crystal of Sect. III.B, but now we shall characterize it without any approximations. Focusing on transverse electric waves with $\mathbf{E} = E_z \hat{\mathbf{z}}$ and $\mathbf{H} = H_x \hat{\mathbf{x}} + H_y \hat{\mathbf{y}}$ , as before, one can easily show that the frequency-domain Maxwell's equations for a system described by the constitutive relations detailed in Sect. III.B can be reduced to:

$$\hat{L} \cdot \mathbf{Q} = \frac{\omega}{c} \mathbf{M} \cdot \mathbf{Q}, \qquad \text{with} \tag{29a}$$

$$\hat{L} = \begin{pmatrix} 0 & -i\partial_y & i\partial_x & -i\frac{\omega_\mathrm{p}(\mathbf{r})}{c} \\ -i\partial_y & 0 & 0 & 0 \\ +i\partial_x & 0 & 0 & 0 \\ i\frac{\omega_\mathrm{p}(\mathbf{r})}{c} & 0 & 0 & 0 \end{pmatrix}, \qquad \text{and} \qquad \mathbf{M} = \begin{pmatrix} 1 & \boldsymbol{\xi}(\mathbf{r})\cdot\hat{\mathbf{x}} & \boldsymbol{\xi}(\mathbf{r})\cdot\hat{\mathbf{y}} & 0 \\ \boldsymbol{\xi}(\mathbf{r})\cdot\hat{\mathbf{x}} & 1 & 0 & 0 \\ \boldsymbol{\xi}(\mathbf{r})\cdot\hat{\mathbf{y}} & 0 & 1 & 0 \\ 0 & 0 & 0 & 1 \end{pmatrix}, \tag{29b}$$

In the above, $\boldsymbol{\xi}(\mathbf{r})$ is the Tellegen vector defined as in Eq. (26), and the state vector is defined by $\mathbf{Q}^\mathrm{T} = \begin{bmatrix} E_z & \eta_0 H_x & \eta_0 H_y & \frac{1}{\varepsilon_0 \omega_\mathrm{p}} j \end{bmatrix}$ with $\eta_0$ the free-space impedance and $j$ the z-component of the electric current density in the metallic background.

The material matrix $\mathbf{M}$ is positive definite provided $|\boldsymbol{\xi}| < 1$. In particular, this requires that $\xi_{\max} = \max|\boldsymbol{\xi}|$ must be smaller than 1, which will be assumed in what follows. Numerical analysis shows that $\xi_{\max}$ is marginally larger than $\xi_0$ in Eq. (26): $\xi_{\max} \approx 1.016\xi_0$. Then, provided $\xi_{\max} < 1$, the spectral problem reduces to $\hat{H} \cdot \Psi = \frac{\omega}{c} \Psi$ with $\hat{H} = \mathbf{M}^{-1/2} \cdot \hat{L} \cdot \mathbf{M}^{-1/2}$ Hermitian and $\Psi = \mathbf{M}^{1/2} \cdot \mathbf{Q}$.

We used the plane wave method to characterize the exact band structure of the dispersive photonic crystal. The calculated band structure is represented in Fig. 4a, for the structural parameters $\xi_0 = 0.67$, $\omega_{\mathrm{p}} a / c = 5.63$ and $R = 0.35a$. Consistent with previous discussion, we see that the spectrum is unbounded with the positive frequencies mirrored with respect to the real axis [Fig. 4a]. Due to this property, the dispersive system lacks of well-defined "ground" with a finite number of frequency branches.

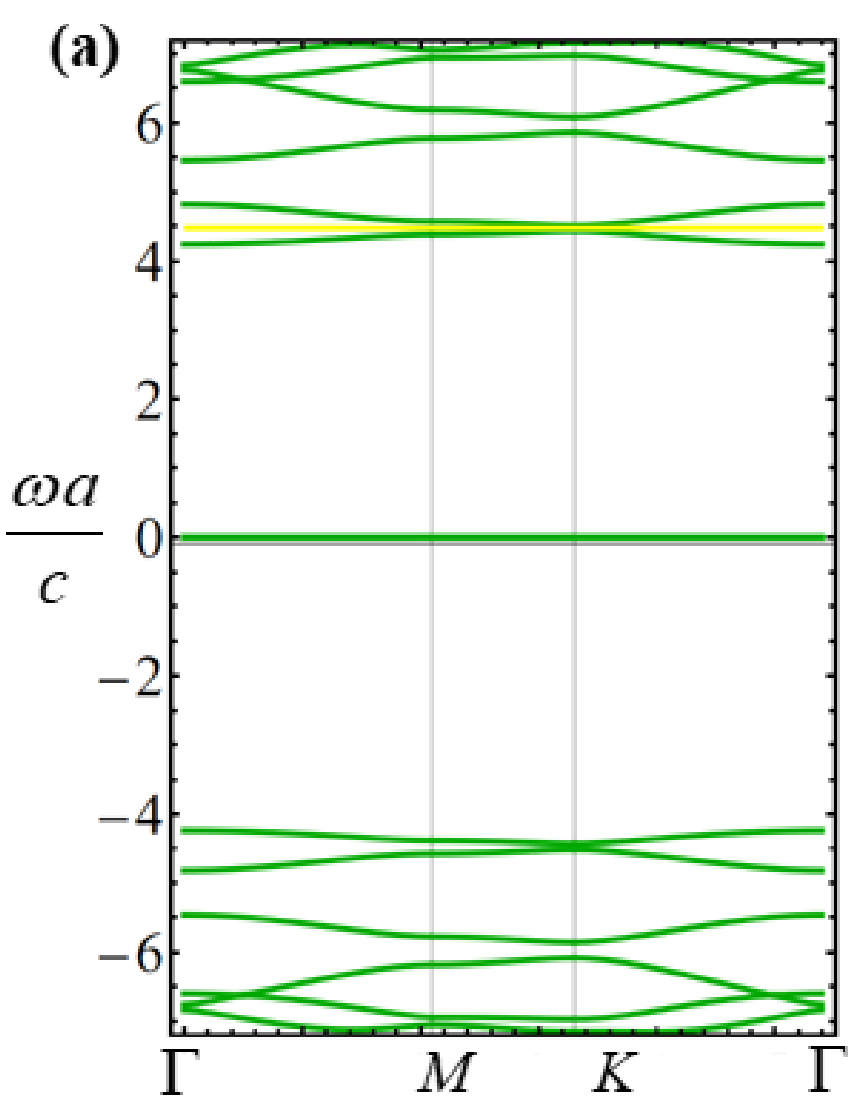

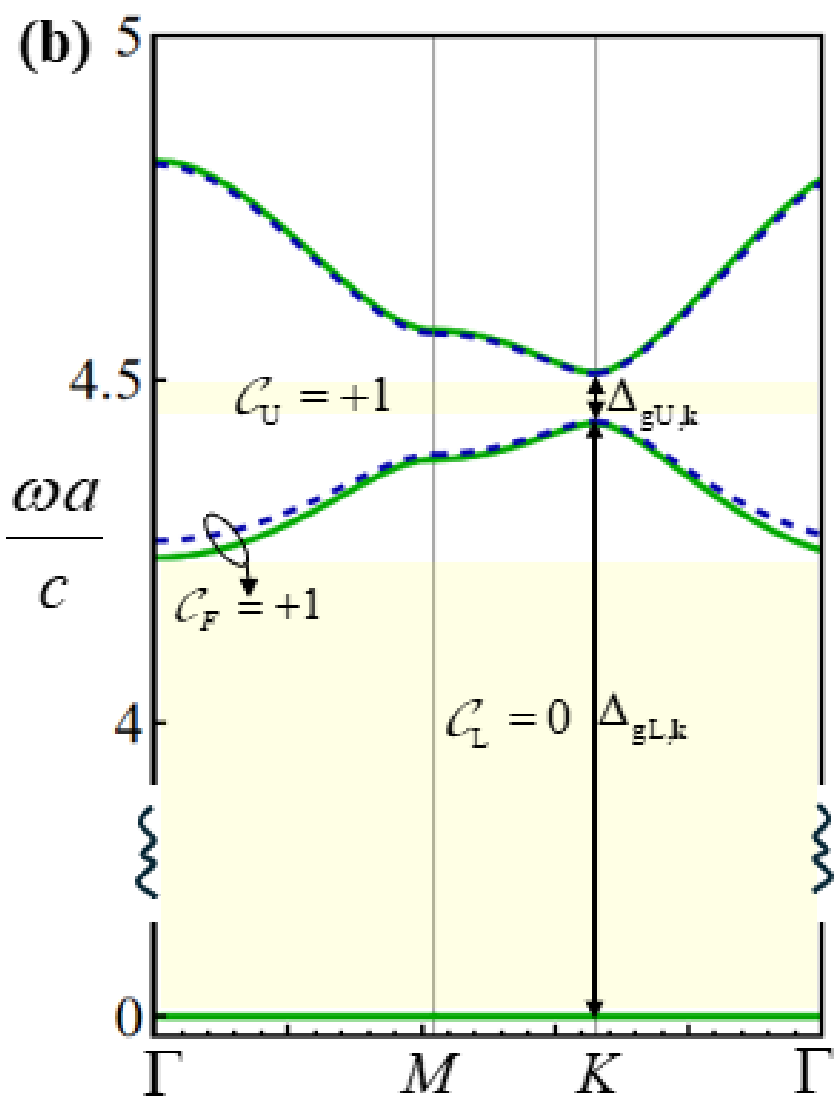

**Fig. 4 (a)** Exact band structure of a dispersive Tellegen-type crystal showing both the positive and negative frequency spectra. The radius of the scatterers is $R = 0.35a$ and the material parameters are $\xi_0 = 0.67$ and $\omega_{\mathrm{p}} a / c = 5.63$. **(b)** Comparison between the exact dispersion of the first few positive-frequency bands (solid green lines) and the dispersion calculated using the approximate scalar formulation of Sect. III.B for $\omega_0 a / c = 4.47$ (dotted blue lines). The gap Chern numbers indicated as $\mathcal{C}_{\mathrm{U}}, \mathcal{C}_{\mathrm{L}}$ are computed for the upper and lower photonic band gaps, with their corresponding direct gap widths indicated by $\Delta_{\mathrm{gU},\mathbf{k}}, \Delta_{\mathrm{gL},\mathbf{k}}$.

It is relevant to compare the exact frequency dispersion with that predicted by the approximate scalar formulation of Sec. III.B. This comparison is shown in Fig. 4b, and reveals good agreement between the two calculation methods near the frequency $\omega_0 a / c = 4.47$ , at which the scalar operator is sampled.

The operator family $\hat{H}_{\mathbf{k}}$ that characterizes the Bloch problem is obtained as usual with the replacement $\nabla \to \nabla + i\mathbf{k}$ in $\hat{H} = \mathbf{M}^{-1/2} \cdot \hat{L} \cdot \mathbf{M}^{-1/2}$ . Clearly, $\hat{H}_{\mathbf{k}}$ depends linearly on $\mathbf{k}$ . Furthermore, its first derivatives with respect to $\mathbf{k}$ are given by:

$$\partial_i \hat{H}_{\mathbf{k}} = \mathbf{M}^{-1/2} \cdot \begin{pmatrix} 0 & \hat{\mathbf{u}}_i \cdot \hat{\mathbf{y}} & -\hat{\mathbf{u}}_i \cdot \hat{\mathbf{x}} & 0 \\ \hat{\mathbf{u}}_i \cdot \hat{\mathbf{y}} & 0 & 0 & 0 \\ -\hat{\mathbf{u}}_i \cdot \hat{\mathbf{x}} & 0 & 0 & 0 \\ 0 & 0 & 0 & 0 \end{pmatrix} \cdot \mathbf{M}^{-1/2}, \qquad (30)$$

where $\hat{\mathbf{u}}_i$ is a unit vector directed along *i*-th direction (for example, $\hat{\mathbf{u}}_1 = \hat{\mathbf{x}}$). It is curious to note that the above formulas – or more generally the linear dependence of $\hat{H}_{\mathbf{k}}$ on $\mathbf{k}$ – imply that $\nabla^2 \hat{H}_{\mathbf{k}} = 0$. Therefore, when the curvature bound in Eq. (13) is applied to a dispersive photonic system, it leads to the indeterminate form $C_{\mathrm{B}} = \frac{1}{\Delta_{\mathrm{g}}} \frac{\pi}{\mathrm{A}_{\mathrm{cell}}} \infty \times 0$. This occurs because the infinite number of bands below any given gap ( $N_{\mathrm{F}} = \infty$ ) is combined with a vanishing curvature norm ($\left\| \nabla_{\mathbf{k}}^2 \hat{H}_{\mathbf{k}} \right\|_{\infty} = 0$).

Even though the curvature bound is not applicable in the dispersive case, one can use the velocity bound (17) to constraint the topological charge of any finite set of bands. In the following, we illustrate this procedure for the first positive frequency band represented in Fig. 4a. To this end, we note that since $\partial_i \hat{H} = \mathbf{M}^{-1/2} \cdot \partial_i \hat{L}_{\mathbf{k}} \cdot \mathbf{M}^{-1/2}$ (here $\partial_i \hat{L}_{\mathbf{k}}$ is the middle matrix in the right-hand side of Eq. (30)), one has $\left\|\partial_i \hat{H}\right\|_\infty \le \left\|\mathbf{M}^{-1/2}\right\|_\infty \left\|\partial_i \hat{L}_{\mathbf{k}}\right\|_\infty \left\|\mathbf{M}^{-1/2}\right\|_\infty = \left\|\mathbf{M}^{-1}\right\|_\infty$. We used $\left\|\partial_i \hat{L}_{\mathbf{k}}\right\|_\infty = 1$ and the fact that for a positive-definite matrix $\left\|\mathbf{M}^{-1/2}\right\|_\infty = \left(\left\|\mathbf{M}^{-1}\right\|_\infty\right)^{1/2}$. Taking into account that $\left\|\mathbf{M}^{-1}\right\|_\infty = 1/\left(1-|\xi|\right)$, we find that $\left\|\partial_i \hat{H}\right\|_\infty \le \max 1/\left(1-|\xi|\right) = 1/\left(1-\xi_{\max}\right)$.

Note that in the present example the velocity bound is not constrained by the (normalized) speed of light in vacuum because the model of the Tellegen material does not respect the constraints of special relativity. Substituting $\left\|\partial_i \hat{H}\right\|_\infty \le 1/\left(1-\xi_{\max}\right)$ in Eq. (17), we obtain the following explicit velocity bound for the topological charge of any subset $F$ of bands of the dispersive Tellegen-type crystal:

$$\left|\mathcal{C}_F\right| \le C_{\mathrm{B}}, \qquad \text{with} \quad C_{\mathrm{B}} = \frac{1}{\Delta_{\mathrm{g}}^2} \frac{4\pi N_{\mathrm{F}}}{\mathrm{A}_{\mathrm{cell}}} \frac{1}{\left(1-\xi_{\max}\right)^2} \tag{31}$$

and $\xi_{\max} \approx 1.016\xi_0$.

We computed the exact topological charge of the dispersive crystal using the Green's-function method of Ref. [31]. We find that the low-frequency topological gap shown in Fig. 4b is trivial, $\mathcal{C}_{\mathrm{L}} = 0$, whereas the topological charge of the second gap is $\mathcal{C}_{\mathrm{U}} = +1$. These results agree with those obtained using the approximate scalar formulation of Sec. III.B. Furthermore, they imply that the Chern number of the first positive frequency band shown in Fig. 4b is $\mathcal{C}_F = \mathcal{C}_{\mathrm{U}} - \mathcal{C}_{\mathrm{L}} = +1$.

We note in passing that the pathologies discussed in Ref. [24], which can lead to ill-defined topological invariants, do not affect the system analyzed here. Indeed, in our crystal, dispersion does not lead to an accumulation of bands at a finite and nonzero frequency for the considered field polarization. The accumulation of bands at zero frequency is harmless, since these bands do not carry Berry curvature. Furthermore, although the system has an infinite number of bands below the gap, the total charge of the negative-frequency bands is necessarily zero, consistent with $\mathcal{C}_{\mathrm{L}} = 0$ . This is because the low-frequency gap never closes when the nonreciprocity parameter $\xi_0$ is turned on, ensuring that the topological charge of that gap remains identical to that of the reciprocal case with $\xi_0 = 0$.

We applied the velocity bound in Eq. (31) to the first positive frequency band of the crystal ( $N_{\mathrm{F}} = 1$ ). In Eq. (31), we use $\Delta_{\mathrm{g}} = \min_{\mathbf{k}} \left\{\Delta_{\mathrm{gU},\mathbf{k}}, \Delta_{\mathrm{gL},\mathbf{k}}\right\}$, with $\Delta_{\mathrm{gU},\mathbf{k}}, \Delta_{\mathrm{gL},\mathbf{k}}$ the lower and upper direct frequency gap widths for a fixed wave vector, as represented in Fig. 4b.

Figure 5 shows the bound $C_{\mathrm{B}}$ as a function of the photonic crystal parameters. Specifically, Fig. 5a corresponds to a sweep of the plasma frequency $\omega_{\mathrm{p}}$, whereas Fig. 5b corresponds to a sweep of the Tellegen-coupling amplitude $\xi_0$. The topological charge of the relevant band remains unchanged throughout these parameter sweeps because the gap width $\Delta_{\mathrm{g}}$ remains strictly positive (not shown).

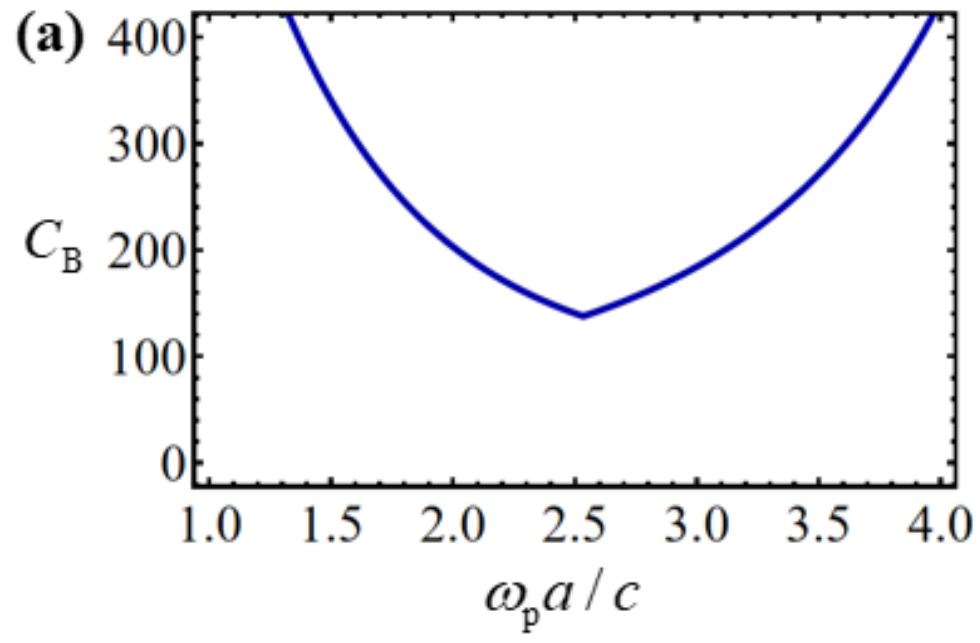

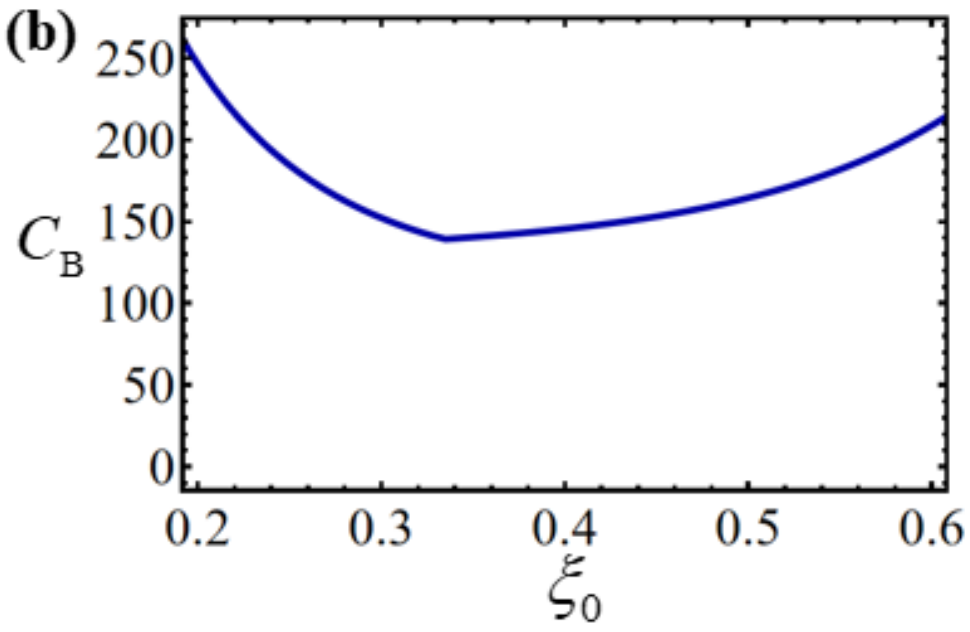

**Fig. 5** Bound on the topological charge $\mathcal{C}_F = \mathcal{C}_{\mathrm{U}} - \mathcal{C}_{\mathrm{L}}$ of the 1st positive frequency band represented in Fig. 4b. **(a)** $C_{\mathrm{B}}$ as a function of the normalized plasma frequency $\omega_{\mathrm{p}} a/c$, for $\xi_0 = 0.35$, and **(b)** $C_{\mathrm{B}}$ as a function of the Tellegen-coupling amplitude $\xi_0$, for $\omega_{\mathrm{p}} a/c = 2.5$. The remaining structural parameters are as in Fig. 4.

Unfortunately, the calculated bound $C_{\mathrm{B}}$ remains consistently quite large, never falling below 140. As the topological charge of the 1st positive frequency band is $\mathcal{C}_F = +1$, the analysis indicates that the velocity bound is loose and fails to meaningfully constrain the band Chern number. We have numerically verified that this behavior is general: it is also observed in other dispersive photonic crystals, such as magnetized plasmas described by the hydrodynamic model (not shown).

We attribute the looseness of the velocity bound to two factors. The first is specific to our crystal: the normalized velocity satisfies $\left\|\partial_i \hat{H}\right\|_\infty \leq 1/\left(1-\xi_{\max}\right)$, meaning that the velocity bound correlates with the strength of the nonreciprocity. As a result, a strong nonreciprocal response, which is beneficial for opening large gaps, penalizes the bound. For example, Fig. 5b shows that the bound $C_{\mathrm{B}}$ becomes very large for large values of $\xi_0$, whereas its actual minimum is attained for $\xi_0 \approx 0.34$.

The second factor is more general. In the velocity bound, the normalized gap is controlled by the square of a frequency difference, $\left(\Delta_{\mathrm{g}} a\right)^2\Big|_{\mathrm{vel}} \sim \left(\frac{\omega_{\max} a}{c} - \frac{\omega_{\min} a}{c}\right)^2$, whereas in the curvature bound it is controlled by a difference of squared frequencies, $\Delta_{\mathrm{g}} a^2\Big|_{\mathrm{curv}} \sim \left(\frac{\omega_{\max} a}{c}\right)^2 - \left(\frac{\omega_{\min} a}{c}\right)^2$. Hence, it follows that $\Delta_{\mathrm{g}} a^2\Big|_{\mathrm{curv}} \sim \left(\Delta_{\mathrm{g}} a\right)^2\Big|_{\mathrm{vel}} \frac{1+\omega_{\max}/\omega_{\min}}{1-\omega_{\max}/\omega_{\min}}$, which shows that the term $\Delta_{\mathrm{g}} a^2\Big|_{\mathrm{curv}}$ can be substantially smaller than $\left(\Delta_{\mathrm{g}} a\right)^2\Big|_{\mathrm{vel}}$, especially for narrow gaps with $\omega_{\max}/\omega_{\min} \sim 1$. This property combined with the fact that the velocity bound is controlled by both the upper and lower direct frequency gaps, explains why the curvature bound, when applicable, can be much tighter than the velocity bound.

### *B. Strategy to use the curvature bound in a dispersive crystal*

Given the limited effectiveness of the velocity bound, it is useful to discuss alternative strategies that can more effectively constrain the gap Chern number of dispersive photonic crystals. Next, we describe a simple strategy applicable to systems such as the Tellegen-type photonic crystal.

Specifically, suppose that a dispersive system is described by some (frequency independent) operator $\hat{H}_{\text{exa}}$ that is topological, and that the electrodynamics of the same system can also be characterized by some frequency-dependent scalar operator $\hat{H}_{\text{sca}}(\omega)$ that is quadratic in the wave vector. Consider some gap of the exact operator and let $\omega_0$ be some frequency lying inside that gap. At frequency $\omega_0$, the exact electromagnetic response of the crystal is described equally well by $\hat{H}_{\text{exa}}$ and $\hat{H}_{\text{sca}}(\omega_0)$. Indeed, at a fixed frequency, the electrodynamics is insensitive to the material dispersion. Evidently, $\hat{H}_{\text{sca}}(\omega_0)$ also supports a band gap at $\omega_0$.

Now the key observation is that the gap Chern number, $\mathcal{C}_{\text{exa}}$, calculated using the exact response $\hat{H}_{\text{exa}}$ in the considered gap, must be exactly coincident with the gap Chern, $\mathcal{C}_{\text{sca}}$, obtained from the approximate scalar operator $\hat{H}_{\text{sca}}(\omega_0)$. The reason is that both $\hat{H}_{\text{exa}}$ and $\hat{H}_{\text{sca}}(\omega_0)$ describe the same electrodynamics at $\omega = \omega_0$, and therefore must predict the same edge states at that frequency when the photonic crystal is paired with a trivial insulator. The direction of propagation of edge states is controlled by the Poynting vector, which is also insensitive to the material dispersion. Consequently, by the bulk-edge correspondence [35, 36], the gap Chern numbers of the two operators (which are properties of the gap, rather than of a particular frequency inside the gap) must be identical: $\mathcal{C}_{\text{exa}} = \mathcal{C}_{\text{sca}}$. As a corollary, any bound on $|\mathcal{C}_{\text{sca}}|$, for example one obtained from the curvature bound, is also a bound on $|\mathcal{C}_{\text{exa}}|$.

To make the discussion concrete consider again the Tellegen crystal described by the operator $\hat{H}_{\text{exa}} = \mathbf{M}^{-1/2} \cdot \hat{L} \cdot \mathbf{M}^{-1/2}$. The very same system can also be described by the scalar operator $\hat{H}_{\text{sca}}(\omega) = -\left(\nabla - i\frac{\omega}{c}\hat{\mathbf{z}} \times \boldsymbol{\xi}(\mathbf{r})\right)^2 + \frac{\omega_{\text{p}}^2(\mathbf{r})}{c^2}$, which, when evaluated at $\omega = \omega_0$, reduces to the operator in Eq. (27). As shown previously, both operators predict essentially the same band structure near $\omega = \omega_0$ in the gap between the $1^{\text{st}}$ and $2^{\text{nd}}$ frequency bands (Fig. 4b), and both operators share the same gap Chern number in that gap $\mathcal{C}_{\text{exa}} = \mathcal{C}_{\text{sca}} = +1$. Therefore, the curvature bounds depicted in Fig. 2, obtained for the approximate scalar Hamiltonian ($|\mathcal{C}_{\text{sca}}| \le C_{\text{B}}$), apply equally well to the gap Chern number of the exact operator, since the two descriptions are guaranteed to share the same gap Chern number. Note that any deformation of $\hat{H}_{\text{sca}}(\omega_0)$ that does not close the gap, can be used to constrain the gap Chern number of the exact operator, even if the corresponding auxiliary system is not directly correlated with $\hat{H}_{\text{exa}}$. This follows from the fact that $\mathcal{C}_{\text{sca}}$ is invariant under gap-preserving deformations. Thus, auxiliary scalar formulations provide a practical route for obtaining meaningful topological bounds in dispersive photonic systems.

## V. Conclusions

In this work, we extended the theory of Onishi and Fu to general bosonic wave systems and derived upper bounds on the gap Chern numbers of photonic crystals. We obtained two types of bounds. The first is a curvature bound, applicable to operator families that are quadratic in the Bloch wave vector, and controlled by the spectral norm of the second-order derivatives of the Bloch operator in momentum space. The second is a velocity bound, applicable to operator families that are linear in the wave vector, and controlled by the spectral norms of the velocity operators. In electromagnetic systems, the latter has a direct relativistic interpretation, since wave propagation is ultimately constrained by the speed of light in the medium.

We showed that the application of the Onishi-Fu result to bosonic systems is not automatic. Unlike fermionic insulators, bosonic systems do not generally have a natural ground-state filling, and the number of bands below a given gap is typically infinite. This difficulty occurs, for example, in dispersive electromagnetic systems, whose spectra contain both positive and negative frequencies, and in vector formulations of Maxwell problems, where longitudinal modes may generate infinitely many flat bands at zero frequency. These features can make curvature-type bounds trivial. The velocity bound partly avoids this difficulty by considering a finite subset of bands and by relying on the boundedness of the velocity operators.

We applied our curvature bound to two representative nondispersive photonic platforms: magnetized gyrotropic photonic crystals and plasmonic photonic crystals with a Tellegen-type magnetoelectric coupling. In both cases, the bound can constrain the gap Chern number using only spectral information, without requiring an explicit topological calculation. We used the invariance of the gap Chern number under gap-preserving deformations to improve the estimates, showing that the curvature bound can be sufficiently tight to constrain the topological charge of the photonic crystals.

By contrast, the application of the velocity bound to a dispersive Tellegen-type photonic crystal led to bounds that were consistently loose. To overcome this limitation, we introduced an alternative strategy based on auxiliary nondispersive operators. This approach enabled us to constrain the topological charge of the exact dispersive system, despite the presence of infinitely many bands below the gap.

Our analysis clarifies both the usefulness and the limitations of universal topological bounds in photonic systems. It shows that meaningful constraints on photonic gap Chern numbers can be obtained, but also that the formulation of the propagation problem is crucial. In particular, avoiding spurious infinite band counts and identifying suitable auxiliary operators are essential for applying such bounds to realistic electromagnetic media.

**Acknowledgements**

This work is supported in part by the Simons Foundation (award SFI-MPS-EWP-00008530-10), by national funds through FCT – Fundação para a Ciência e a Tecnologia, I.P., and, when eligible, co-funded by EU funds under project/support UID/50008/2025 – Instituto de Telecomunicações, with DOI identifier https://doi.org/10.54499/UID/50008/2025, and by Grant No. ID2021-129035NB-I00 funded by MCIN/AEI/10.13039/501100011033 and by ERDF/EU.

## Appendix A: The f-sum rule

For completeness, in this appendix we provide a self-contained proof of the f-sum rule (11) [23]. To this end, we define

$$
\begin{aligned}
F_i &= \sum_{\substack{n\mathbf{k}\in F \\ m\mathbf{k}\in E}} \left\langle u_{n\mathbf{k}} \left| \hat{v}_{i,\mathbf{k}} \right| u_{m\mathbf{k}} \right\rangle \left\langle u_{m\mathbf{k}} \left| \hat{v}_{i,\mathbf{k}} \right| u_{n\mathbf{k}} \right\rangle \frac{1}{E_{mn,\mathbf{k}}} . \\
&= \sum_{\substack{n\mathbf{k}\in F \\ m\mathbf{k}\in E}} \left\langle \partial_i u_{n\mathbf{k}} \left| u_{m\mathbf{k}} \right\rangle \left\langle u_{m\mathbf{k}} \right| \partial_i u_{n\mathbf{k}} \right\rangle E_{mn,\mathbf{k}} .
\end{aligned} \tag{A1}
$$

In the second identity, we used

$$
\left\langle u_{m\mathbf{k}} \left| \partial_i \hat{H}_{\mathbf{k}} \right| u_{n\mathbf{k}} \right\rangle = -E_{mn,\mathbf{k}} \left\langle u_{m\mathbf{k}} \mid \partial_i u_{n\mathbf{k}} \right\rangle + \delta_{mn} \partial_i E_{n\mathbf{k}} . \tag{A2}
$$

This is obtained by differentiating $\left(\hat{H}_{\mathbf{k}} - E_{n\mathbf{k}}\right)\left|u_{n\mathbf{k}}\right\rangle = 0$ with respect to the wave vector,

$$
\left(\partial_i \hat{H}_{\mathbf{k}} - \partial_i E_{n\mathbf{k}}\right)\left|u_{n\mathbf{k}}\right\rangle = -\left(\hat{H}_{\mathbf{k}} - E_{n\mathbf{k}}\right)\left|\partial_i u_{n\mathbf{k}}\right\rangle , \tag{A3}
$$

and applying the bra $\left\langle u_{m\mathbf{k}}\right|$ to both sides of the equation.

As the generic term of summation in Eq. (A1) is odd under an interchange of the indices *m* and *n*, the summation over *m* can be unconstrained and replaced by a summation over all modes. Then, using the spectral decomposition of $\hat{H}_{\mathbf{k}} = \sum_m E_{m\mathbf{k}} \left|u_{m\mathbf{k}}\right\rangle\left\langle u_{m\mathbf{k}}\right|$, we can rewrite (A1) as:

$$
F_i = \sum_{n\mathbf{k}\in F} \left\langle \partial_i u_{n\mathbf{k}} \left| \left(\hat{H}_{\mathbf{k}} - E_{nk}\right) \right| \partial_i u_{n\mathbf{k}} \right\rangle . \tag{A4}
$$

To proceed, we differentiate both sides of $\partial_i E_{n\mathbf{k}} = \left\langle u_n \mid \partial_i \hat{H}_{\mathbf{k}} \left| u_{n\mathbf{k}} \right\rangle\right.$ [Eq. (A2) with *m*=*n*], to obtain $\partial_i \partial_j E_{n\mathbf{k}} = \left\langle u_{n\mathbf{k}} \mid \partial_i \partial_j \hat{H}_{\mathbf{k}} \left| u_{n\mathbf{k}} \right\rangle\right. + \left\langle \partial_j u_{n\mathbf{k}} \mid \partial_i \hat{H}_{\mathbf{k}} \left| u_{n\mathbf{k}} \right\rangle\right. + \left\langle u_{n\mathbf{k}} \mid \partial_i \hat{H}_{\mathbf{k}} \left| \partial_j u_{n\mathbf{k}} \right\rangle\right.$, which can be rewritten as:

$$
\left\langle u_{n\mathbf{k}} \mid \partial_i \partial_j \hat{H}_{\mathbf{k}} - \partial_i \partial_j E_{n\mathbf{k}} \left| u_{n\mathbf{k}} \right\rangle\right. = -2\,\mathrm{Re}\left\{\left\langle \partial_j u_{n\mathbf{k}} \mid \partial_i \hat{H}_{\mathbf{k}} \left| u_{n\mathbf{k}} \right\rangle\right.\right\} . \tag{A5}
$$

Combining this result with Eq. (A3) and noting that $\left\langle \partial_j u_{n\mathbf{k}} \mid u_{n\mathbf{k}} \right\rangle$ is pure imaginary, we obtain the key identity:

$$\left\langle u_{n\mathbf{k}} \,|\, \partial_i \partial_j \hat{H}_\mathbf{k} - \partial_i \partial_j E_{n\mathbf{k}} \left| u_{n\mathbf{k}} \right\rangle = 2\,\mathrm{Re}\left\{ \left\langle \partial_j u_{n\mathbf{k}} \,|\left( \hat{H}_\mathbf{k} - E_{nk} \right)\right| \partial_i u_{n\mathbf{k}} \right\rangle \right\}. \tag{A6}$$

Substituting the above formula with $i = j$ in Eq. (A4), we find that [23]:

$$F_i = \frac{1}{2} \left\langle u_{n\mathbf{k}} \left| \partial_i^2 \hat{H}_\mathbf{k} - \partial_i^2 E_{n\mathbf{k}} \right| u_{n\mathbf{k}} \right\rangle. \tag{A7}$$

Hence, by integrating $F_i$ in Eq. (A1) over the Brillouin zone, we obtain the desired result [Eq. (11)]. Note that for periodic systems the integral of $\partial_i^2 E_{n\mathbf{k}}$ vanishes, $\int_{\mathrm{BZ}} d^2\mathbf{k}\, \partial_i^2 E_{n\mathbf{k}} = 0$, because $E_{n\mathbf{k}}$ and its derivatives in the wave vector are periodic functions. As a side comment, we note that this term cannot not be dropped in continuum systems (i.e., with no intrinsic periodicity, see for example Ref. [26]), because in such systems the contribution of $\nabla E_{n\mathbf{k}}$ at the boundary of the integration domain ($k = \infty$) does not need to vanish.

## Appendix B: Velocity bound for non-Hermitian operators

In this Appendix, we generalize the velocity bound (17) to the case of non-Hermitian operators [29, 30]. In the non-Hermitian case ($\hat{H}_\mathbf{k} \neq \hat{H}_\mathbf{k}^\dagger$), the relevant spectral problem takes the form:

$$\hat{H}_\mathbf{k} \left| u_{n\mathbf{k}}^{\mathrm{R}} \right\rangle = E_{n\mathbf{k}} \left| u_{n\mathbf{k}}^{\mathrm{R}} \right\rangle, \qquad \hat{H}_\mathbf{k}^\dagger \left| u_{n\mathbf{k}}^{\mathrm{L}} \right\rangle = E_{n\mathbf{k}}^* \left| u_{n\mathbf{k}}^{\mathrm{L}} \right\rangle. \tag{B1}$$

Here, $\left| u_{n\mathbf{k}}^{\mathrm{R}} \right\rangle$ and $\left| u_{n\mathbf{k}}^{\mathrm{L}} \right\rangle$ are the "right" and "left" eigenfunctions of the operator. The eigenvalues $E_{n\mathbf{k}}$ are generally complex. We shall suppose that $\hat{H}_\mathbf{k}$ is diagonalizable. In such case, it is possible to choose the left and right eigenfunctions such that they form a bi-orthogonal basis [29]:

$$\left\langle u_{n\mathbf{k}}^{\mathrm{L}} \,|\, u_{m\mathbf{k}}^{\mathrm{R}} \right\rangle = \delta_{nm}. \tag{B2}$$

The gap Chern number of a subset of bands $F$ separated from the rest of the spectrum (complementary subset $E$) is given by (compare with Eq. (2)) [29, 30]:

$$\begin{aligned}\mathcal{C}_F &= \mathrm{Re}\left\{\frac{i}{2\pi}\int d^2\mathbf{k}\sum_{n\mathbf{k}\in F}\left[\left\langle\partial_1 u_{n\mathbf{k}}^{\mathrm{L}}\middle|\partial_2 u_{n\mathbf{k}}^{\mathrm{R}}\right\rangle-\left\langle\partial_2 u_{n\mathbf{k}}^{\mathrm{L}}\middle|\partial_1 u_{n\mathbf{k}}^{\mathrm{R}}\right\rangle\right]\right\}\\ &= \mathrm{Re}\left\{\frac{i}{2\pi}\int d^2\mathbf{k}\sum_{\substack{n\mathbf{k}\in F\\ m\mathbf{k}\in E}}\left[\left\langle u_{n\mathbf{k}}^{\mathrm{L}}\middle|\partial_1\hat{H}_\mathbf{k}\middle|u_{m\mathbf{k}}^{\mathrm{R}}\right\rangle\left\langle u_{m\mathbf{k}}^{\mathrm{L}}\middle|\partial_2\hat{H}_\mathbf{k}\middle|u_{n\mathbf{k}}^{\mathrm{R}}\right\rangle-1\leftrightarrow 2\right]\frac{1}{\left[E_{mn,\mathbf{k}}\right]^2}\right\}\end{aligned}\tag{B3}$$

The term $1\leftrightarrow 2$ is identical to the first term, except that the subscripts of the velocity operators are interchanged. The second identity is obtained by using the completeness relation $\sum_{m\mathbf{k}}\left|u_{m\mathbf{k}}^{\mathrm{R}}\right\rangle\left\langle u_{m\mathbf{k}}^{\mathrm{L}}\right|$, and a generalization of the relation derived in Appendix A: $E_{mn,\mathbf{k}}\left\langle u_{m\mathbf{k}}^{\mathrm{L}}\middle|\partial_i u_{n\mathbf{k}}^{\mathrm{R}}\right\rangle+\left\langle u_{m\mathbf{k}}^{\mathrm{L}}\middle|\partial_i\hat{H}_\mathbf{k}\middle|u_{n\mathbf{k}}^{\mathrm{R}}\right\rangle=0$ for $m\neq n$.

Similar to the main text, we can rewrite Eq. (B3) as:

$$\mathcal{C}_F = \mathrm{Re}\left\{\frac{i}{2\pi}\sum_{n\mathbf{k}\in F}\int d^2\mathbf{k}\left\langle u_{n\mathbf{k}}^{\mathrm{L}}\middle|\mathbf{A}_{n\mathbf{k}}^{12}\middle|u_{n\mathbf{k}}^{\mathrm{R}}\right\rangle-\left\langle u_{n\mathbf{k}}^{\mathrm{L}}\middle|\mathbf{A}_{n\mathbf{k}}^{21}\middle|u_{n\mathbf{k}}^{\mathrm{R}}\right\rangle\right\}.\tag{B4}$$

where $\mathbf{A}_{n\mathbf{k}}^{ij}=\partial_i\hat{H}_\mathbf{k}G_{\mathbf{k},n}\partial_j\hat{H}_\mathbf{k}$ with $G_{\mathbf{k},n}=\sum_{m\mathbf{k}\in E}\frac{1}{\left[E_{mn,\mathbf{k}}\right]^2}\left|u_{m\mathbf{k}}^{\mathrm{R}}\right\rangle\left\langle u_{m\mathbf{k}}^{\mathrm{L}}\right|$. Next, we use:

$$\left|\left\langle u_{n\mathbf{k}}^{\mathrm{L}}\middle|\mathbf{A}_{n\mathbf{k}}^{ij}\middle|u_{n\mathbf{k}}^{\mathrm{R}}\right\rangle\right|\leq\left\|u_{n\mathbf{k}}^{\mathrm{L}}\right\|\left\|\mathbf{A}_{n\mathbf{k}}^{ij}\left|u_{n\mathbf{k}}^{\mathrm{R}}\right\rangle\right\|\leq\left\|u_{n\mathbf{k}}^{\mathrm{L}}\right\|\left\|u_{n\mathbf{k}}^{\mathrm{R}}\right\|\left\|\mathbf{A}_{n\mathbf{k}}^{ij}\right\|_\infty.\tag{B5}$$

Note that $\mathbf{A}_{n\mathbf{k}}^{ij}$ is typically a non-Hermitian operator. The spectral norm of a non-Hermitian operator $\mathbf{B}$ is determined by $\|\mathbf{B}\|_\infty=\left(\max|\lambda_i|^2\right)^{1/2}$, with $|\lambda_i|^2$ the eigenvalues of the (Hermitian) non-negative operator $\mathbf{B}^\dagger\mathbf{B}$. Similar to the Hermitian case, we have $\|\mathbf{B}_1\mathbf{B}_2...\mathbf{B}_N\|_\infty\leq\|\mathbf{B}_1\|_\infty\|\mathbf{B}_2\|_\infty...\|\mathbf{B}_N\|_\infty$. In addition, we have the useful relation $\|\mathbf{B}\|_\infty=\left\|\mathbf{B}^\dagger\right\|_\infty$ [28].

From the above considerations, we immediately obtain the bound:

$$|\mathcal{C}_F| \le \frac{2}{2\pi} \sum_{n\mathbf{k}\in F} \int d^2\mathbf{k} \left\| u_{n\mathbf{k}}^{\mathrm{L}} \right\| \left\| u_{n\mathbf{k}}^{\mathrm{R}} \right\| \left\| \partial_1 \hat{H}_{\mathbf{k}} \right\|_\infty \left\| \partial_2 \hat{H}_{\mathbf{k}} \right\|_\infty \left\| G_{\mathbf{k},n} \right\|_\infty . \tag{B6}$$

Note that different from the Hermitian case, $\left\| u_{n\mathbf{k}}^{\mathrm{L}} \right\| \left\| u_{n\mathbf{k}}^{\mathrm{R}} \right\|$ does not need to be identical to one, and its square determines the so-called Petermann factor [37].

In order to characterize $\left\| G_{\mathbf{k},n} \right\|_\infty$, it is useful to identify the operator $\hat{H}_{\mathbf{k}}$ with a matrix, and the vectors $\left| u_{m\mathbf{k}}^{\mathrm{R}} \right\rangle$ with the columns of a matrix $\mathbf{R}_{\mathbf{k}}$ such that $\hat{H}_{\mathbf{k}} \mathbf{R}_{\mathbf{k}} = \mathbf{R}_{\mathbf{k}} \cdot \mathbf{\Delta}_{\mathbf{k}}$, where $\mathbf{\Delta}_{\mathbf{k}}$ is a diagonal matrix with the eigenvalues $E_{m\mathbf{k}}$. Then, it follows that $\hat{H}_{\mathbf{k}}^\dagger \mathbf{L}_{\mathbf{k}} = \mathbf{L}_{\mathbf{k}} \cdot \mathbf{\Delta}_{\mathbf{k}}^*$, with $\mathbf{L}_{\mathbf{k}} = \left( \mathbf{R}_{\mathbf{k}}^{-1} \right)^\dagger$. Thus, the vectors $\left| u_{m\mathbf{k}}^{\mathrm{L}} \right\rangle$ can be identified with the columns of $\mathbf{L}_{\mathbf{k}} = \left( \mathbf{R}_{\mathbf{k}}^{-1} \right)^\dagger$.

It can now be checked that $G_{\mathbf{k},n} = \sum_{m\mathbf{k}\in E} \frac{1}{\left[ E_{mn,\mathbf{k}} \right]^2} \left| u_{m\mathbf{k}}^{\mathrm{R}} \right\rangle \left\langle u_{m\mathbf{k}}^{\mathrm{L}} \right|$ can be identified with:

$$G_{\mathbf{k},n} = \tilde{\mathbf{R}}_{\mathbf{k}} \mathbf{D}_{\mathbf{k},n} \tilde{\mathbf{L}}_{\mathbf{k}}^\dagger . \tag{B7}$$

Here, $\mathbf{D}_{\mathbf{k},n}$ is a diagonal square matrix with elements $1/\left[ E_{mn,\mathbf{k}} \right]^2$, $\tilde{\mathbf{R}}_{\mathbf{k}}$ and $\tilde{\mathbf{L}}_{\mathbf{k}}$ are rectangular matrices obtained from $\mathbf{R}_{\mathbf{k}}$ and $\mathbf{L}_{\mathbf{k}}$, respectively, by dropping all the columns in subset *F*. From the properties of the spectral norm discussed previously, it is clear that $\left\| G_{\mathbf{k},n} \right\|_\infty \le \left\| \tilde{\mathbf{R}}_{\mathbf{k}} \right\|_\infty \left\| \mathbf{D}_{\mathbf{k},n} \right\|_\infty \left\| \tilde{\mathbf{L}}_{\mathbf{k}} \right\|_\infty \le \frac{1}{\Delta_{\mathrm{g}}^2} \left\| \tilde{\mathbf{R}}_{\mathbf{k}} \right\|_\infty \left\| \tilde{\mathbf{L}}_{\mathbf{k}} \right\|_\infty$, with $\Delta_{\mathrm{g}}$ the minimum direct spectral gap [Eq. (9)]. Note that the spectral gap is evaluated with the complex spectrum.

The spectral norm is such that $\left\| \mathbf{B} \right\|_\infty = \max \left\| \mathbf{B} \cdot \mathbf{x} \right\|$ with $\left\| \mathbf{x} \right\| = 1$ [28]. Using this property, one can see that $\left\| \tilde{\mathbf{R}}_{\mathbf{k}} \right\|_\infty \le \left\| \mathbf{R}_{\mathbf{k}} \right\|_\infty$ and $\left\| \tilde{\mathbf{L}}_{\mathbf{k}} \right\|_\infty \le \left\| \mathbf{L}_{\mathbf{k}} \right\|_\infty$ (recall that $\tilde{\mathbf{R}}_{\mathbf{k}}$ is formed by a subset of columns

of $\mathbf{R_k}$ ). Furthermore, as $\left\|u_{n\mathbf{k}}^{\mathrm{R}}\right\|$ can be identified with the spectral norm of a matrix whose single column is $u_{n\mathbf{k}}^{\mathrm{R}}$ , we also have $\left\|u_{n\mathbf{k}}^{\mathrm{R}}\right\| \leq \left\|\mathbf{R_k}\right\|_\infty$ , and likewise $\left\|u_{n\mathbf{k}}^{\mathrm{L}}\right\| \leq \left\|\mathbf{L_k}\right\|_\infty$ .

The previous discussion shows that $\left\|G_{\mathbf{k},n}\right\|_\infty \leq \frac{1}{\Delta_{\mathrm{g}}^2}\kappa_{\mathbf{k}}$ and $\left\|u_{n\mathbf{k}}^{\mathrm{R}}\right\|\left\|u_{n\mathbf{k}}^{\mathrm{L}}\right\| \leq \kappa_{\mathbf{k}}$ with $\kappa_{\mathbf{k}} = \left\|\mathbf{R_k}\right\|_\infty \left\|\mathbf{L_k}\right\|_\infty = \left\|\mathbf{R_k}\right\|_\infty \left\|\mathbf{R_k^{-1}}\right\|_\infty$ the *condition number* of the eigenvectors matrix. Note that the condition number can be seen as a generalization of the Petermann factor. In the Hermitian case, $\mathbf{R_k}$ is a unitary matrix ( $\mathbf{R_k^\dagger R_k} = \mathbf{1}$ ), and thus its spectral norm is exactly 1, so that $\kappa_{\mathbf{k}} = 1$ . Thus, the condition number measures the deviation from the unitary condition.

Substituting $\left\|G_{\mathbf{k},n}\right\|_\infty \leq \frac{1}{\Delta_{\mathrm{g}}^2}\kappa_{\mathbf{k}}$ and $\left\|u_{n\mathbf{k}}^{\mathrm{R}}\right\|\left\|u_{n\mathbf{k}}^{\mathrm{L}}\right\| \leq \kappa_{\mathbf{k}}$ into Eq. (B3), and proceeding as in the main text, we readily find that

$$\left|\mathcal{C}_F\right| \leq \frac{4\pi N_{\mathrm{F}}}{A_{\mathrm{cell}}\Delta_{\mathrm{g}}^2}\kappa_{\max}^2 \max\left\|\partial_1 \hat{H}_{\mathbf{k}}\right\|_\infty \left\|\partial_2 \hat{H}_{\mathbf{k}}\right\|_\infty , \tag{B8}$$

with $\kappa_{\max} = \max \kappa_{\mathbf{k}}$ . This is the generalization of the velocity bound [Eq. (17)] to non-Hermitian systems.